\documentclass[useAMS,usenatbib]{mn2e} 
\usepackage{graphicx}
\usepackage{amsmath}

\title[Star Formation in the Shapley Supercluster]{ACCESS II: A Complete Census of Star Formation in the Shapley Supercluster -- UV and IR Luminosity Functions}

\author[Haines et al.]{C. P. Haines,$^{1}$ G. Busarello,$^2$ P. Merluzzi,$^2$ R. J. Smith,$^{3}$ S. Raychaudhury,$^{1}$ \and A. Mercurio,$^2$ G. P. Smith$^{1}$\\
$^{1}$School of Physics and Astronomy, University of Birmingham, Edgbaston, Birmingham, B15 2TT, UK; cph@star.sr.bham.ac.uk\\
$^{2}$INAF - Osservatorio Astronomico di Capodimonte, via Moiariello 16, I-80131 Napoli, Italy\\
$^{3}$Department of Physics, University of Durham, Durham DH1 3LE, UK\\
}

\begin{document}

\maketitle
\label{firstpage}

\begin{abstract}
We present panoramic {\em Spitzer}/MIPS mid- and far-infrared (MIR/FIR) and {\em GALEX} ultraviolet imaging of the the most massive and dynamically active system in the local Universe, the Shapley supercluster at $z{=}0.048$, covering the five clusters which make up the supercluster core. We combine these data with existing spectroscopic data from 814 confirmed supercluster members to produce the first study of a local rich cluster including both ultraviolet and infrared luminosity functions (LFs). This joint analysis allows us to produce a complete census of star-formation (both obscured and unobscured), extending down to ${\rm SFRs}{\sim}0.$02--0.$05\,{\rm M}_{\odot}{\rm yr}^{-1}$, and quantify the level of obscuration of star formation among cluster galaxies, providing a local benchmark for comparison to ongoing and future studies of cluster galaxies at higher redshifts with {\em Spitzer} and {\em Herschel}.
 The {\em GALEX} near-ultraviolet (NUV) and far-ultraviolet (FUV) luminosity functions (LFs) obtained have steeper faint-end slopes than the local field population, due largely to the contribution of massive, quiescent galaxies at $M_{FUV}{\ga}{-}16$. The 24$\mu$m and 70$\mu$m galaxy LFs for the Shapley supercluster instead have shapes fully consistent with those obtained for the Coma cluster and for the local field galaxy population. This apparent lack of environmental dependence for the shape of the FIR luminosity function suggests that the bulk of the star-forming galaxies that make up the observed cluster infrared LF have been recently accreted from the field and have yet to have their star formation activity significantly affected by the cluster environment.
We estimate a global SFR of $327\,{\rm M}_{\odot}{\rm yr}^{-1}$ over the whole supercluster core, of which just ${\sim}2$0 per cent is visible directly in the ultraviolet continuum and ${\sim}8$0 per cent is reprocessed by dust and emitted in the infrared. The level of obscuration ($L_{IR}/L_{FUV}$) in star-forming galaxies is seen to increase linearly with $L_{K}$ over two orders of magnitude in stellar mass.

\end{abstract}

\begin{keywords}
galaxies: active --- galaxies: clusters: general --- galaxies: evolution --- galaxies: stellar content --- galaxies: clusters: individual (A3558) --- galaxies: clusters: individual (A3562) --- galaxies: clusters: individual (A3556)
\end{keywords}

\section{Introduction}
\label{intro}

\setcounter{footnote}{3}

The conversion of gas into stars is one of the most fundamental astrophysical processes regulating galaxy evolution, and hence the star formation rate (SFR) which measures the rate at which this conversion occurs, building up the stellar mass of the galaxy, is a key observable in galaxy evolution studies. The harsh cluster environment is well known to affect the star formation activity of the member galaxies, as quantified through the star-formation (SF)--density relation \citep[e.g.][]{dressler85,balogh00,lewis,gomez}. While various physical mechanisms such as ram-pressure stripping, harassment and starvation have been proposed to quench star formation in infalling spiral galaxies transforming them into the passive lenticulars which dominate cluster cores \citep[for reviews see e.g.][]{boselli06,haines07}, the dominant evolutionary pathway(s) remains unresolved.

Many of these processes that drive the evolution of galaxies also shape the luminosity function (LF), one of the most basic and fundamental properties of the galaxy population \citep{benson}. The LFs of cluster galaxies at various wavelengths can hence provide quantitative probes of how these dense environments affect the fundamental galaxy properties such as the overall mass function, stellar masses and SFRs. A key question is whether the LF shows a significant environmantal dependence or is instead universal. This remains unclear, with some studies suggesting relatively little variation with environment \citep[e.g.][]{depropris,christlein,rines}, while other find sigificant differences including brighter characterisitic luminosities and steeper faint-end slopes \citep[e.g.][]{popesso,sos1}, which could be largely ascribed to the diverse morphological composition of cluster and field populations \citep{delapparent}. 
However the optical luminosities of galaxies depend not only on their stellar masses but are also strongly affected by the presence of young stellar populations and dust, which can bias the optical LFs, making it difficult to reliably interpret the data. The near-infrared (NIR) instead is much less sensitive to the effects of dust or star formation, and hence the NIR (e.g. $K$-band) LF can be considered a reliable estimator of the underlying stellar mass function \citep{belldejong}. 
The ultraviolet (${\sim}2$\,000{\AA}) instead provides a measure of the recent SFR over the last $10^{8}$\,yr in galaxies, being dominated by emission from young stars of intermediate masses \citep[2--5\,M$_{\odot}$;][]{boselli}, and hence the ultraviolet LF represents a useful tool to quantify the effects of the cluster environment on star formation. 

Ultraviolet radiation from young stars can be strongly attenuated by the intervening dust, and indeed the ultraviolet emission from spiral galaxies comes predominately from those slightly older stars ($10^{7}{-}10^{8}$yrs) that have migrated away from their birthplaces in the highly obscured giant molecular clouds \citep{calzetti05}. The amount of attenuation depends both on the general properties of the galaxy (mass, metallicity, morphology) and the relative geometry of the dust and young stars, but for many galaxies more than 90\% of the UV photons are absorbed by dust, corresponding to ${\rm A(FUV)}{>}2.5$ and hence potentially signficantly biasing the ultraviolet LF. This energy absorbed by the dust is however reprocessed as thermal radiation in the mid/far-infrared (8--1000$\mu$m), and so observations at these wavelengths allow us to quantify this energy and thus infer the amount of obscured star formation \citep[e.g.][]{kennicutt,calzetti07,rieke09,kennicutt09}.
Optically-thin infrared and radio continuum emission provides an inherently extinction-free estimate of SFRs, and so the infrared and radio LFs can be considered complementary to the ultraviolet LFs, being sensitive to obscured emission missed by the ultraviolet LF, and between them provide strong constraints on the global SFRs in cluster galaxies and on the impact of cluster related environmental processes on star formation.  

The first pioneering ultraviolet observations of cluster galaxies were performed using the sounding rocket payloads of \citet{smith82}. 
The FOCA balloon-borne 40cm-diameter telescope surveyed the local Coma and Abell 1367 clusters \citep{donas90,donas91}, obtaining the 2000{\AA} ultraviolet luminosity function for Coma. They found that despite Coma being the archetype of the elliptical-rich and concentrated cluster, in the ultraviolet it is strongly dominated by spirals, and its surface density shows low concentration and contrast over the field. \citet{cortese03} then combined the FOCA and later FAUST UV cluster surveys to produce a composite UV luminosity function for the Virgo, Coma and Abell 1367 clusters. This was found to be well fitted by a Schechter function with M$^{*}_{UV}{=}{-}18.79{\pm}0.40$ and $\alpha{=}{-}1.50{\pm}0.10$ values that did not differ significantly from the local UV luminosity function of the field. More recent deeper studies of the UV LFs of Coma and Abell 1367 with {\em GALEX} showed steeper faint-end slopes than for the field, due to the contribution of passively-evolving galaxies at faint magnitudes \citep{cortese05,cortese08}.

The {\em Infrared Astronomical Satellite} \citep[{\em IRAS};][]{iras} provided the first opportunity to study the infrared properties of cluster galaxies and to compare them with the field population. In a pointed {\em IRAS} survey of the Hercules cluster at $z{=}0.036$ reaching 50mJy at 60$\mu$m ($L_{IR}{\sim}5{\times}10^{9}L_{\odot}$), \citet{young} identified twenty-four 60$\mu$m sources as spiral galaxies belonging to the cluster, producing a luminosity function that could be fit by a double power law. Notably galaxies classified as ellipticals or S0s were absent from the infrared data. {\em ISO} and more recently the {\em Spitzer} space telescope \citep{werner} have allowed much more sensitive infrared studies of star formation in clusters, revealing ubiquitous obscured star formation among cluster galaxies whose contribution was underestimated by ${\ga}10{\times}$ by previous optical studies \citep{duc,metcalfe}.
For local clusters the resultant SFRs remain low \citep[${\la}10\,{\rm M}_{\odot}\,{\rm yr}^{-1}$;][]{duc,wolf,haines10} and the MIR luminosity functions are consistent with those found in field regions \citep{bai06,bai}. At higher redshifts, galaxies with obscured SFRs of 30--100\,M$_{\odot}\,{\rm yr}^{-1}$ become commonplace, indicative of rapid evolution, but largely paralleling that evolution found in the field \citep{zheng,haines09,vulcani}, consistent with these galaxies representing a recently accreted population \citep{geach06,geach09,marcillac,koyama}.

\begin{figure*}
\centerline{\includegraphics[width=170mm]{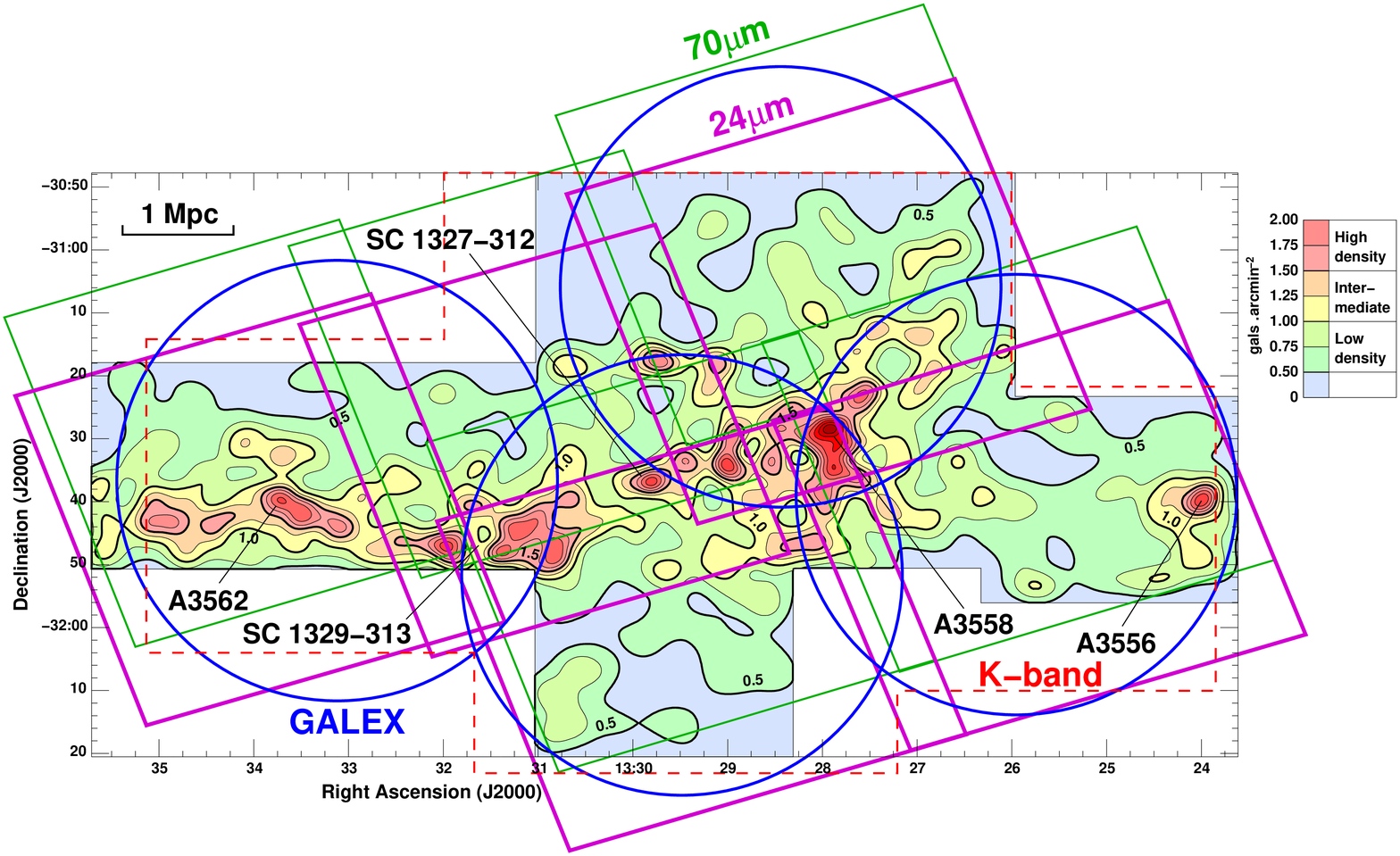}}
\caption{The multi-wavelength photometric coverage of the ACCESS survey. The shaded region indicates the WFI $B,R$-band imaging of the Shapley Optical Survey. Coloured contours represent the surface density of $R{<}21$ galaxies (Fig.~1 of \citet{sos2}). The respective coverages of the {\em GALEX} FUV/NUV (blue circles), WFCAM/$K$-band (red dashed-line), {\em Spitzer} 24$\mu$m (magenta lines) and 70$\mu$m (green lines) imaging are all indicated.}
\label{map}
\end{figure*}

In \citet[Paper I]{merluzzi} we introduced the ACCESS multi-wavelength survey of the Shapley supercluster core (SSC) as well as presented an analysis of the $K$-band data. 
The target has been chosen since the most dramatic effects of environment on galaxy evolution should occur in superclusters, where the infall and encounter velocities of galaxies are greatest (${>}1$\,000\,km\,s$^{-1}$), groups and clusters are still merging, and significant numbers of galaxies will be encountering the dense intra-cluster medium (ICM) of the supercluster environment for the first time. Moreover, the interplay of a variety of physical processes need to be invoked to describe the overdensity of baryons on supercluster scales \citep[e.g.][]{fabian91}, which emphasises the cosmological context of studies involving the effects of star formation, AGN and associated gas processes.

The Shapley supercluster was identified as the largest overdensity of galaxies and clusters in the $z{<}0.1$ Universe in an APM survey of galaxies covering the entire southern sky \citep{sr89} and confirmed by early X-ray studies \citep{sr91,day91}.  In the original Abell catalogue \citep{aco89}, the core of the Shapley supercluster was found to be represented by a single cluster (A3558$\equiv$Shapley~8) of optical richness~4, inconsistent with the observed velocity dispersion of the cluster. Early X-ray observations from {\em Einstein} and {\em ROSAT} \citep{breen94a} resolved this by discovering that the core in fact comprises of a chain three merging clusters, A3558, SC1327-312 and SC1329-313. The SSC in this paper refers to these three clusters, along with the two associated Abell clusters on either side, A3562 and A3556 \citep[see][for a general description]{sos1}.

In this paper we attempt to build a {\em complete} and {\em unbiased} census of star formation in galaxies belonging to the SSC using a multi-wavelength analysis combining {\em GALEX} ultraviolet and {\em Spitzer} infrared observations, in order to circumvent any biases that may exist in previous surveys in which star formation rates in cluster galaxies were estimated from a single SFR indicator. 
By assembling a comprehensive pan-chromatic (FUV--FIR) dataset for the SSC we are able to fully account for both obscured and unobscured star-formation, and by dealing with {\em global} photometric measurements, we are independent of the aperture biases \citep{kewley} prevalent in many previous studies based on optical spectra obtained from fibre-fed spectrographs \citep[e.g.][]{lewis,gomez,haines07}.
 In {\S}~\ref{sec:data} we present the ultraviolet and new infrared datasets and describe their reduction and resultant source catalogues.
In {\S}~\ref{uvlf} and {\S}~\ref{mirlf} we present the ultraviolet (NUV/FUV) and infrared (24$\mu$m, 70$\mu$m) galaxy luminosity functions for the Shapley supercluster. We then perform a joint analysis of the ultraviolet and infrared data to measure their relative contributions to the global supercluster SFR budget in {\S}~\ref{sec:uvir}, and quantify the level of obsuration among cluster galaxies (both individually and globally), before presenting the discussion in {\S}~\ref{sec:discuss}. 

In a companion paper \citep[Paper III]{paper3} we examine in detail the {\em nature} of star formation in galaxies within the Shapley supercluster, finding a robust bimodality in the $f_{24}/f_{K}$ colours that can be understood as the well-known split into star-forming and passive galaxies, as well as taking advantage of available 1.4\,GHz VLA radio continuum data from \citet{miller} to show via examination of the panchromatic (FUV--FIR) SEDs that the bulk of the global supercluster SFR comes from quiescent star formation in normal infalling spiral disks who have yet to be affected by the supercluster environment.
In a further follow-up paper (Haines et al. 2010c, in preparation) we will examine the environmental trends in star formation, and identify and quantify various classes of galaxies in the process of being transformed from star-forming spiral to passive S0. 

This work is carried out in the framework of the joint research programme ACCESS\footnote{{\it A Complete Census of Star-formation and nuclear activity in the Shapley supercluster}, PI: P. Merluzzi, a collaboration among the Universities of Durham and Birmingham (UK), the Italian National Institute of Astrophysics' Osservatorio di Capodimonte and the Australian National University (ANU).} \citep[http://www.na.astro.it/ACCESS]{merluzzi} aimed at distinguishing among the mechanisms which drive galaxy evolution across different ranges of mass by their interactions with the environment. 
When necessary we assume \mbox{$\Omega_{M}{=}0.3$}, \mbox{$\Omega_{\Lambda}{=}0.7$} and \mbox{H$_{0}{=}70\,$km\,s$^{-1}$Mpc$^{-1}$}, such that at the distance of the Shapley supercluster 1\,arcsec is equivalent to 0.96\,kpc.
All magnitudes are quoted in the Vega system unless otherwise stated.

\section{Data}
\label{sec:data}

The ultraviolet and mid-infrared data analysed in this paper are complemented by existing panoramic optical $B$- and $R$-band imaging from the Shapley Optical Survey \citep[SOS:][]{sos1,sos2} and near-infrared $K$-band imaging \citep{merluzzi}, which cover the same region. Figure~\ref{map} shows the Shapley supercluster core (indicated by isodensity contours) with superposed the relative coverages of the different aspects of our multi-wavelength survey. The ultraviolet (blue circles) and mid-infrared (24$\mu$m magenta; 70$\mu$m green) photometry cover essentially the same regions, the bulk of which are also covered by the $K$-band imaging (red dashed-lines), while the earlier SOS covers a slightly narrower region. 
We thus have full multi-wavelength coverage of the main filamentary structure connecting A3562 and A3558, and the extension to A3556.  

\subsection{The Shapley Optical Survey}

The SOS comprises wide-field optical imaging of a 2.2\,deg$^{2}$ region covering the whole of the Shapley supercluster core. The observations were made from March 2002 to April 2003 using the Wide Field Imager (WFI) camera, on the 2.2m MPG/ESO telescope at La Silla. 
The SOS is made up of eight contiguous WFI fields of $34{\times}33$\,arcmin$^{2}$ (pixel scales of 0.238\,arcsec), each with total exposure times of 1500s in $B$ and 1200s in $R$, and typical FWHMs of 0.7--1.0\,arcsec, resulting in galaxy catalogues that are both complete and reliable to $R{=}22.0$ and $B{=}22.5$. Full details of the observations, data reduction, and the production of the galaxy catalogue are described in \citet{sos1}.
The high quality of the optical images has allowed morphological classifications and structural parameters to be derived for all supercluster galaxies to ${\sim}M_{R}^{*}{+}3$ \citep{gargiulo} within the automated {\sc 2dphot} environment \citep{labarbera}.

\subsection{$K$-band and spectroscopic data}

The $K$-band survey of the SSC was carried out with the Wide Field infrared CAMera (WFCAM) on the 3.8m United Kingdom Infra-Red Telescope (UKIRT) in April 2007 as part of the UKIRT service programme (P.I. R.J. Smith). The WFCAM data consist of five complete $52{\times}52$arcmin$^2$ tiles (made up of four interleaved multiframe exposures), covering a region of 3.0\,deg$^2$, with exposure times of 300s, 0.4\,arcec pixel scale, FWHMs of 0.9--1.2\,arcsec, and reach $K{=}19.5$ at $5{\sigma}$. For more details see \citet{merluzzi}.

The Shapley supercluster core has been observed by a variety of redshift surveys \citep{quintana,bardelli98,bardelli00,kaldare,drinkwater,smith07,cava,6df} resulting in 964 galaxies in our $K$-band catalogue having redshift information, of which 814 lie in the velocity range of the SSC ($10\,700{<}cz{<}17\,000$\,km\,s$^{-1}$). Of these, 415 are from the AAOmega-based spectroscopic survey of \citet{smith07}. We have redshifts for all 107 $K{<}12.3$ ($K{<}K^{*}{+}0.6$) galaxies within the $K$-band survey. The 90\% and 50\% spectroscopic completeness limits are $K{=}13.25$ ($K^{*}{+}1.5$) and $K{=}14.3$ ($K^{*}{+}2.5$) respectively.

\subsection{The GALEX observations}
 
\begin{figure}
\centerline{\includegraphics[width=60mm]{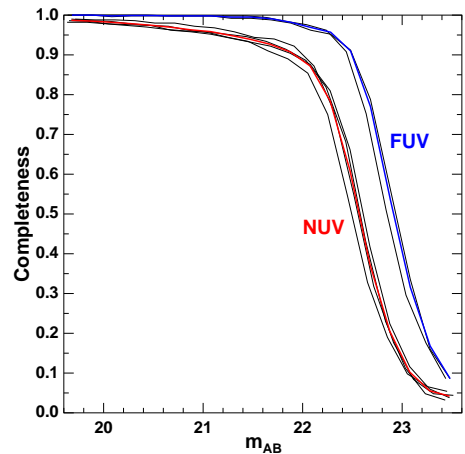}}
\caption{The completeness of the {\em GALEX} NUV (thick red) and FUV (thick blue) photometric catalogues as a function of $m_{AB}$, averaged over the four {\em GALEX} images. The thin black lines indicate the completeness levels for the individual {\em GALEX} NUV and FUV images.}
\label{uv_completeness}
\end{figure}

The UV data analysed in this paper consist of four contiguous {\em GALEX} pointings (blue circles in Fig.~\ref{map}) making up the Cycle 4 Guest Investigation proposal GI4-098 (P.I. R.J. Smith), and which were observed in May 2008. The {\em GALEX} field of view is circular with diameter 1.2$^{\circ}$, with pixel scale 1.5\,arcsec. Each pointing is simulaneously observed in both the far-UV (FUV) and near-UV (NUV) bands with effective wavelengths of 1516{\AA} and 2267{\AA}, and spatial resolutions of 4.3 and 5.3\,arcsec respectively. The exposure times are in the range 1238--1688\,s, i.e. comparable to the {\em GALEX} Medium Imaging Survey. 
See \citet{martin05} and \citet{morrissey05,morrissey07} for details regarding the {\em GALEX} instruments and mission.

Sources were detected and measured from the {\em GALEX} images using {\sc SExtractor} \citep{bertin}. As the NUV images are deeper than the FUV images (in terms of source counts), we used the NUV images for detection and measured the FUV flux in the same aperture as for the NUV. Following \citet{sos1} and \citet{gray}, we adopted an optimized, dual (`hot' and `cold') configuration, in order to deblend high-surface brightness objects that are close on the sky in projection, and simultaneously avoid the spurious shredding of highly structured spiral galaxies into multiple star-forming regions. 
The FUV and NUV magnitudes are corrected for Galactic extinction using the \citet{schlegel} reddening map and the parametrization of the Galactic extinction law given by \citet{cardelli} for a total-to-selective extinction ratio of $R_{V}{=}3.1$, resulting in the conversion factors $A_{FUV}{=}7.9\,E(B-V)$ and $A_{NUV}{=}8.0\,E(B-V)$. For our fields $E(B-V)=0.0$48--0.058, resulting in corrections of ${\sim}0.40$\,mag. 

The flux detection limits and completeness of each {\em GALEX} image were determined by individually inserting 500 simulated sources at random positions across the {\em GALEX} image for a range of fluxes within a given magnitude bin, and determining their detection rate and recovered magnitudes, using identical extraction procedures. This is then repeated for all magnitude bins of interest. As for all but the brightest galaxies, sources are unresolved in the {\em GALEX} images, we used the NUV and FUV point-spread functions provided by the {\em GALEX} website as our simulated sources. The resulting completeness levels as a function of $m_{AB}$ for the NUV (red curve) and FUV (blue curve) over the range $19.5{<}m_{AB}{<}23.5$ are shown in Fig.~\ref{uv_completeness}. We expect our UV catalogues to be 90\% complete to $m_{AB}(NUV){=}22.0$ and $m_{AB}(FUV){=}22.5$. Based on the ultraviolet component of the combined FUV+24$\mu$m calibration (Eq.~\ref{leroy}) of \citet{leroy}, which itself comes from the FUV--SFR relation of \citet{salim07} the FUV completeness limit corresponds to a SFR of 0.014\,M$_{\odot}$yr$^{-1}$ at the SSC distance.

\begin{figure}
\centerline{\includegraphics[height=70mm,angle=-90]{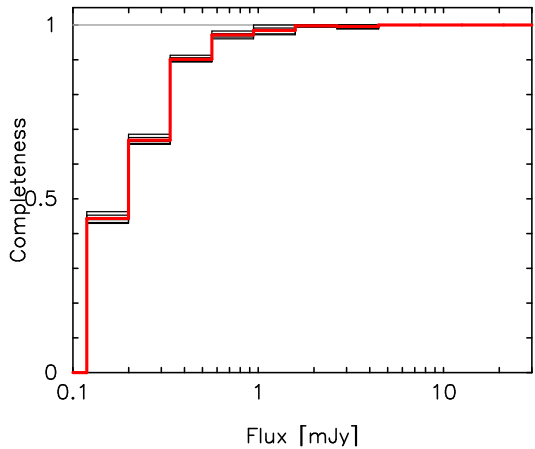}}
\caption{Completeness function of our 24$\mu$m catalogue. Each one of the five black histograms corresponds to the mean of the two simulations for each field. The red histogram is the average over the whole survey area. According to these simulations, the survey is 90\% complete to 0.35mJy.}
\label{compl_24um}
\end{figure}  

The UV detections were cross-matched with the optical-NIR catalogues using a search radius of 5\,arcsec (except for two large edge-on spirals for which a search radius of 15\,arcsec was required). For the remainder of this article, we limit our analysis of the {\em GALEX} data to either the 2.83\,deg$^{2}$ region also covered by $K$-band photometry, or the 2.26\,deg$^{2}$ region also covered by the SOS $B$- and $R$-band photometry.

\subsection{The Spitzer observations}

The panoramic {\em Spitzer} mid-infrared observations of the Shapley supercluster core were carried out over 27--30 August 2008 within the Cycle 5 GO programme 50510 (PI: C.P. Haines).  
The observations consist of five contiguous mosaics observed with MIPS \citep{rieke} in medium scan mode, with coverages at 24$\mu$m and 70$\mu$m shown as magenta and green boxes respectively in Figure~\ref{map}. This scan rate leads to 10 exposures each of 4\,sec, for each pixel along a single scan. As the usable 70$\mu$m array only covers half the full MIPS array width, we used a scan leg spacing equal to the half-array width (160\,arcsec). This strategy results in a homogeneous coverage across both 24$\mu$m and 70$\mu$m mosaics, with each point covered by two 24$\mu$m scans and one 70$\mu$m scan during a single observation.

The reduction of the MIPS data was performed with the MOPEX package \citep[version 18.1.5]{makmar} and with the Germanium Reprocessing Tools \citep[GeRT, version 060415;][]{gordon}\footnote{This software, as well as full information on data analysis is available at the Spitzer Science Center website {\em http://ssc.spitzer.caltech.edu/dataanalysistools/}.}.

\subsubsection{24$\mu$m data}
\label{mopex}

We reduced each of the five fields (corresponding to the five AORs) separately, with the aim to allow for cross-checking during the reduction. Since the mosaics delivered by the pipeline were affected by dark latents along the scan direction, we started by applying to the Basic Calibrated Data (BCD) the so-called `self-calibration', consisting of dividing the BCDs by a flat-field derived from the normalised median of all BCDs in each AOR (using the MOPEX module {\em flatfield.pl}). We then produced a mosaic for each field using the MOPEX module {\em mosaic.pl} \citep{makkhan}. The average effective exposure time per pixel in the final mosaics is 84s.

\begin{figure}
\centerline{\includegraphics[width=80mm]{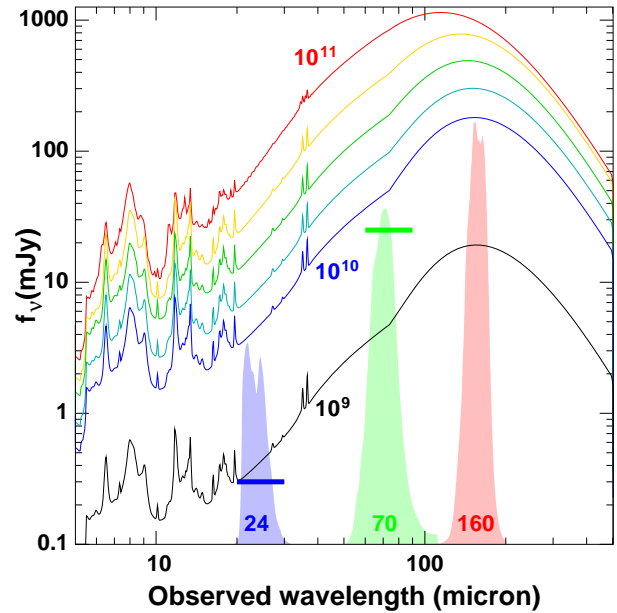}}
\caption{Comparison of the {\em Spitzer} survey depths at 24$\mu$m (blue bar) and 70$\mu$m (green bar) to the luminosity-dependent infrared SEDs of \citet{rieke09} redshifted to the Shapley supercluster. The response functions of the {\em Spitzer}/MIPS filters are indicated by the filled curves.}
\label{rieke}
\end{figure}  

For the detection of the sources, MOPEX relies on an empirical determination of the Point Response Function (PRF) from the actual data. To derive the PRFs for our mosaics, we first produced a theoretical PRF using STinyTim, a program for computing PRF models starting from instrumental parameters. The model PRF was then used to extract a first catalogue of point sources. From the 30--40 brightest sources of each field, we selected ${\sim}10$ `bona-fide' point sources. These were then used to derive the PRF with the MOPEX module {\it prf\_estimate.pl}. The catalogues were extracted using the MOPEX module {\it apex\_1frame.pl}, which, besides source detection, also derives fluxes by fitting the PRF to the detected sources. The fit is performed inside a normalisation radius, which we fixed to 5.9\,arcsec, which includes the central, brightest part of the PRF.
In total 25\,718 sources were detected with SNR$>$3, which was the minimum signal-to-noise ratio we set for a potentially real detection \citep[c.f.][]{fadda06}\footnote{Notice that our SNR values are estimated with a different algorithm of MOPEX, not yet available at the time of \citet{fadda06}. The new SNR values should be doubled to be compared with \citet{fadda06}}. 

To assess the reliability of source extraction and flux estimates and to determine the completeness of our catalogues, we repeated exactly the same process as outlined above on a set of simulated images. To create the simulated images, we first subtracted from the original mosaics the catalogued sources, using the MOPEX module {\em apex\_qa.pl}. On these `empty' fields, we then added the same point sources of the original catalogues, in random positions, but avoiding the areas previously occupied by the `real' sources, where residuals of the subtraction could be present. To create the new point sources, we multiplied our PRFs by the fluxes in the catalogues. This process was conceived in order to put the artificial (but realistic) point sources on as realistic a background as possible. We repeated the process twice on each of the five fields. The simulations were performed using {\sc fortran} codes plus the {\em apex\_qa.pl} module. 

\begin{figure}
\centerline{\includegraphics[height=70mm,angle=-90]{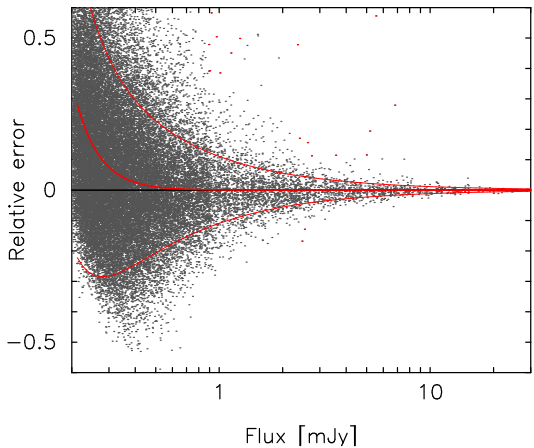}}
\caption{Relative error of the measured 24$\mu$m flux as a function of the input flux, as derived from simulations. Points correspond to the simulated point sources. The thicker red curve corresponds to the mean error, and the top and bottom curves to ${\pm}1{\sigma}$ from the mean error. These curves are used to estimate our flux measurement errors. Mean errors increase for the lower fluxes because faint sources are preferentially detected on the top of positive noise fluctuations, as already noticed by \citet{fadda06}.} 
\label{reliability_24um}
\end{figure}

We computed the completeness function by dividing, for each flux bin, the number of detected sources by the number of input sources. The completeness functions of our 24$\mu$m data are given in Fig.~\ref{compl_24um}, where we show the results for the five fields and their average, from which it is apparent that the completeness functions are fully consistent among each other. From this figure we conclude that our 24$\mu$m data are 90\% complete to a flux of 0.35\,mJy, which we adopt as our survey limit. In Figure~\ref{rieke} we show how this limit relate to the luminosity-dependent infrared SEDs of \citet{rieke09} placed at the distance of the Shapley supercluster. From this figure we see the 24$\mu$m limit corresponds to $\log L_{IR}({\rm erg\,s}^{-1}){=}42.46$ or $L_{IR}{=}7.5{\times}10^{8}L_{\odot}$. Based on the infrared component of the combined FUV+24$\mu$m calibration of \citet{leroy}, or equivalently the 24$\mu$m calibration of \citet{rieke09}, this corresponds to an obscured SFR of 0.05\,M$_{\odot}$yr$^{-1}$, i.e. comparable, but marginally higher than the FUV-based limit.

Fig.~\ref{reliability_24um} shows the relative error of the measured flux as a function of the `true' (input) flux as derived from simulations. Each point corresponds to a point source, while the red curves correspond to the mean error (the central, thicker curve) and to the 1-$\sigma$ limits (top and bottom curves). These last curves were used to estimate our flux measurement errors. These errors could then added to the uncertainty of ${\sim}4$\% on the absolute flux calibration at 24$\mu$m (see the MIPS Instrument Handbook at the Spitzer Science Center web site) and to the 2\% uncertainty in the aperture corrections (see below) to obtain the total uncertainty on the fluxes.

\begin{figure}
\centerline{\includegraphics[width=60mm]{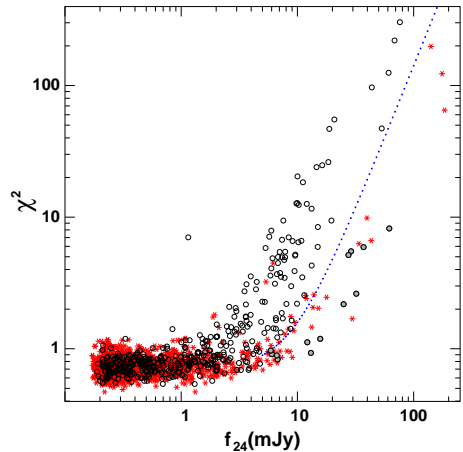}}
\caption{The relation between the $\chi^{2}$ goodness-of-fit value of the MIPS point response function to the 24$\mu$m emission and the total 24$\mu$m flux for Shapley supercluster members (open symbols) and stars (red stars). The blue dotted curve indicates the lower limit for the $\chi^{2}$ distribution of 24$\mu$m supercluster galaxies clearly resolved by MIPS. Those supercluster galaxies whose 24$\mu$m emission is unresolved are indicated by shaded symbols.} 
\label{chi2}
\end{figure}

Since the fluxes were estimated inside a radius 5.9 arcsec, we must apply aperture corrections to obtain the total fluxes from the sources. To this purpose, we derived growth curves for a number of different cases: (i) model point sources with black-body temperatures of 15K, 30K, 50K, 500K (created with STinyTim); (ii) the growth curve given in the MIPS Data Handbook (Fig. 3.2 in version 3.3.1); (iii) the growth curves given by \citet{fadda06}; (iv) the growth curve in the Spitzer Science Center web site. We also measured the growth curve from one bright point source at the centre of a field, which gave a result consistent with the others, but with a small systematic positive offset, which we ascribe to incomplete subtraction of the surrounding sources. All of these growth curves agree, to within 2\%, to determine an aperture correction of $1.69{\pm}0.03$ for our 24$\mu$m fluxes.

In total we associate 4663 $f_{24}{>}0.35$mJy sources with galaxies in the $K$-band image, of which 490 have confirmed redshifts, 394 of which place them in the Shapley supercluster. We have redshifts for all $f_{24}{>}26$mJy sources, and 50\% redshift completeness at ${\sim}5$\,mJy. 

\subsubsection{24$\mu$m photometry of extended sources}

The vast majority of sources detected in our 24$\mu$m mosaics are unresolved at the spatial resolution of the MIPS instrument (FWHM$\sim5^{\prime\prime}$), and so the optimal estimate of their 24$\mu$m fluxes should be obtained via the point-spread function fitting approach of MOPEX. However, at the relatively low redshift of the Shapley supercluster, the brightest galaxies have radii somewhat larger (5--20$^{\prime\prime}$) than the MIPS point-spread function, and can be seen to be resolved by the 24$\mu$m imaging. In these cases, the sources are poorly fitted by point-source functions in the 24$\mu$m images, as witnessed by the large residuals left after subtraction of the best-fit point source from the images, and the resultant fluxes produced by MOPEX will be systematically underestimated. 

\begin{figure}
\centerline{\includegraphics[height=70mm,angle=-90]{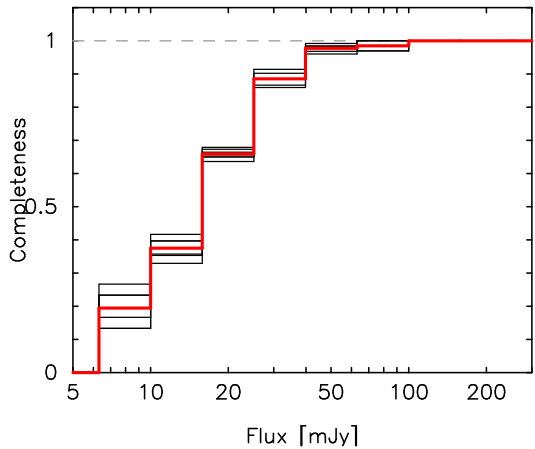}}
\caption{Completeness function of our 70$\mu$m catalogue. The six black histograms correspond to the results of the six simulations, while the red histogram is their average. According to these simulations, the survey is 90\% complete to 25mJy.}
\label{compl_70um}
\end{figure}

We can separate those galaxies resolved by MIPS, and those for which the galaxies appear as point-sources, using the $\chi^{2}$ goodness-of-fit value produced by MOPEX when fitting the 24$\mu$m emission by the MIPS point spread function. Figure~\ref{chi2} shows the relation between this $\chi^{2}$ goodness-of-fit value and the total 24$\mu$m flux for supercluster galaxies (open symbols) and stars (red stars). For $f_{24}{\ga}5$\,mJy we can distinguish two sequences, one corresponding to galaxies and a second with lower $\chi^{2}$ values due to point sources of stellar origin. Along this second sequence, we also find a small population of supercluster galaxies (shaded circles) whose $\chi^{2}$ values are closer to those of stars having the same 24$\mu$m flux levels, indicating that their 24$\mu$m emission is unresolved by MIPS. 

To deal with these extended sources, we also applied SExtractor \citep{bertin} to the median-subtracted 24$\mu$m mosaics (fixing the sky background level in SExtractor to zero), performing photometry in a range of seven circular apertures with diameters in the range 10--90\,arcsec, and correcting the apertures fluxes based on growth curves obtained in {\S}~\ref{mopex}.
For bright galaxies ($R{<}15$ or $K{<}12.5$) and those galaxies for which the 24$\mu$m photometry was poorly fit by a point-spread function within MOPEX, we then compared the flux estimate within the 30\,arcsec diameter aperture obtained by SExtractor with the MOPEX estimate, prefering the SExtractor-based flux estimate in cases where it was found to be ${>}3{\sigma}$ higher than that from MOPEX. If this were the case, we then considered the larger apertures in turn, again preferring the flux estimate from the larger aperture if it were found to be ${>}3{\sigma}$ higher than that of the smaller aperture. 
In total, of the 25\,718 sources detected by MOPEX in the 24$\mu$m mosaics, we identified 231 sources for which the aperture-corrected {\sc SExtractor}-based flux estimate was used instead of the MOPEX one. However, the fraction of galaxies for which the 24$\mu$m flux estimate is {\sc SExtractor}-based increases with 24$\mu$m flux, such that they represent the majority of galaxies with $f_{24}{>}2.5$mJy.

\subsubsection{70$\mu$m data}

\begin{table*}
\begin{tabular}{cccccccccc} \hline
RA & Dec & cz & $K$ & $R$ & $B-R$ & $f_{24}$ & $f_{70}$ & FUV & NUV \\
(J2000) & (J2000) & (km/s) & (mag) & (mag) & & (mJy) & (mJy) & (AB mag) & (AB mag) \\ \hline
13:23:28.949 & -31:43:12.96 & 14036 & --- & 18.845 & 0.447 & $0.313{\pm}0.107$ & $0.0{\pm}0.0$ & 20.290 &20.355\\
13:23:43.990 & -31:47:00.55 & 14325 & --- & 16.540 & 0.875 & $1.518{\pm}0.111$ & $0.0{\pm}0.0$ & 19.084 & 19.726\\
13:23:49.351 & -31:42:56.89 & 12922 & --- & 16.698 & 0.987 & $2.902{\pm}0.112$ & $19.8{\pm}6.4$ & 21.808 & 24.114\\
13:24:05.597 & -31:36:36.11 & 14711 & 16.561 & 18.952 & 1.145 & $0.000{\pm}0.000$ & $0.0{\pm}0.0$ & 99.000 & 99.000\\ 
13:24:20.441 & -31:23:31.05 & 13434 & 15.203 & --- & --- & $2.478{\pm}0.112$ & $34.0{\pm}6.6$ & 20.160 & 20.966\\ \hline
\end{tabular}
\caption{Spectroscopically confirmed supercluster members covered by either our $K$-band or optical imaging. Non-detections in a given passband are indicated by a magnitude of 99 or a flux of zero. No value is given in the case when a galaxy is not covered by a particular passband. The full version of this table is given in the electronic version of the journal - see Supporting information. A portion is shown here for guidance regarding its form and content.} 
\label{catalogue}
\end{table*}

The brightest sources in the 70$\mu$m images were affected by negative side-lobes due to the online filtering when processing of the BCDs with the automated pipeline. To solve this problem, we performed a double-pass filtering using the Germanium Reprocessing Tools. The double-pass filtering consists of an initial filtering along the columns of the BCDs, followed by a high-pass time median filtering, avoiding the pixels near to bright sources, which for this purpose are previously masked. This process minimizes data artifacts (allowing us to eliminate the side-lobes) while preserving the calibration \citep{frayer}.

\begin{figure}
\centerline{\includegraphics[height=70mm,angle=-90]{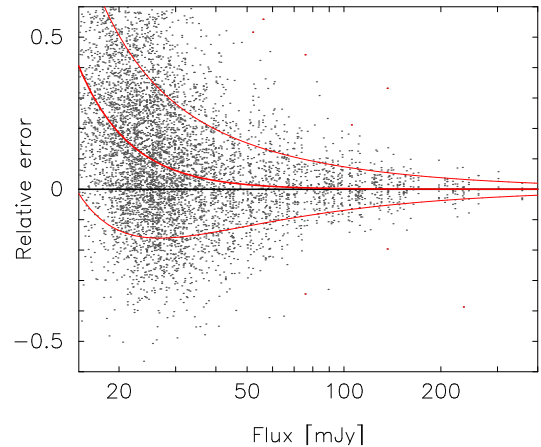}}
\caption{Relative error of the measured 70$\mu$m flux as a function of the input flux, as derived from simulations. Points corresponds to the simulated point sources. The thicker red curve corresponds to the mean error, and the top and bottom curves to $\pm$ 1$\sigma$ from the mean error. These curve are used to estimate our flux measurement errors.}
\label{reliability_70um}
\end{figure}

A single mosaic was created starting from all of the $\sim$9700 filtered BCDs. The subsequent processing was analogous to the 24$\mu$m data. The average exposure time per pixel on the final mosaic turned out to be 42s. The normalisation radius for the PRF fitting was in this case 17 arcsec, corresponding to the beginning of the first Airy minimum for the 70$\mu$m PRF. A total of 1740 sources were selects with SNR$>$3. Simulated images were created in the same way as for the 24$\mu$m data, except that in this case we created six images of the same field. The resulting completeness function is given in Fig.~\ref{compl_70um}, which shows that we reach the 90\% completeness limit at 25 mJy, which we adopt as our survey limit. Based on the infrared SEDs of \citet{rieke09}, this corresponds to $\log(L_{IR}/L_{\odot}){=}9.76$ or $\log(L_{IR}[{\rm erg\,s}^{-1}]){=}43.34$ (Fig.~\ref{rieke}). Based on the direct 70$\mu$m SFR calibration of \citet{calzetti10}, ${\rm SFR}_{70{\mu}{\rm m}}{=}L_{70}({\rm erg\,s}^{-1})/1.7{\times}10^{43}$, this corresponds to a SFR of 0.4\,M$_{\odot}$yr$^{-1}$, a factor 10 less sensitive than the 24$\mu$m limit. Alternatively, the 
infrared SFR calibration of \citet{buat08}, SFR$_{IR}{\rm M}_{\odot}{\rm yr}^{-1}{=}(1{-}{\eta})10^{-9.97}(L_{IR}/L_{\odot})$, produces an identical SFR limit in the case when ${\eta}=0.3$ \citep[following][]{iglesias}, where $\eta$ is the fraction of dust emission due to heating from evolved stars.

Figure~\ref{reliability_70um} shows the relative flux errors as a function of flux, along with the curves encompassing 1$\sigma$ deviation from the mean error, which at our survey limit is 6.5\,mJy. To this error we added another 7\% due to the absolute flux calibration at 70$\mu$m (MIPS Instrument Handbook) and the 3\% uncertainty in the aperture corrections (see below) to obtain the total uncertainty on the fluxes.

As for the 24$\mu$m data, to derive the aperture corrections we analysed different growth curves, in particular: (i) four model point-sources, created with STinyTim, with black-body temperatures of 15K, 30K, 50K and 500K; (ii) three growth curves given by the Spitzer Science Center web-site (two black-bodies at 15K and 3000K and one power-law spectrum). From these curves we derived an aperture correction of 1.74, consistent within 3\% among all curves.

We identify 728 sources with $f_{70}{>}25$mJy, of which 589 are covered by our $K$-band imaging. Of these, 173 have no $K$-band counterpart within 5\,arcsec, but we expect most of these to be either asteroids or high-redshift ULIRGs/QSOs, the former lacking $K$-band counterparts due to their movement across the sky in the time between the two sets of observations. A further 35 sources were classified as stars in the $K$-band image, but we note that almost all of these are probably high-redshift AGN rather than stars given their high $f_{70}/f_{K}$ ratios. In total, we associate 381 $f_{70}{>}25$mJy sources with galaxies in the $K$-band image, of which 194 have confirmed redshifts, 160 of which place them in the Shapley supercluster. 

In Table~\ref{catalogue} we present the positions, redshifts, UV, optical and NIR magnitudes, and {\em Spitzer}/MIPS fluxes of all galaxies spectroscopically confirmed as belonging to the SSC covered by either our $K$-band or optical imaging.

\section{UV luminosity functions}
\label{uvlf}

To determine the UV luminosity function of galaxies in a cluster a reliable measure of the contribution from back/foreground galaxies to the UV counts is required. Where spectroscopic information is lacking, we must rely on statistical methods to estimate the background contamination per magnitude bin, and/or use the UV-optical colours to separate supercluster and background galaxies. 

\begin{figure*}
\centerline{\includegraphics[width=150mm]{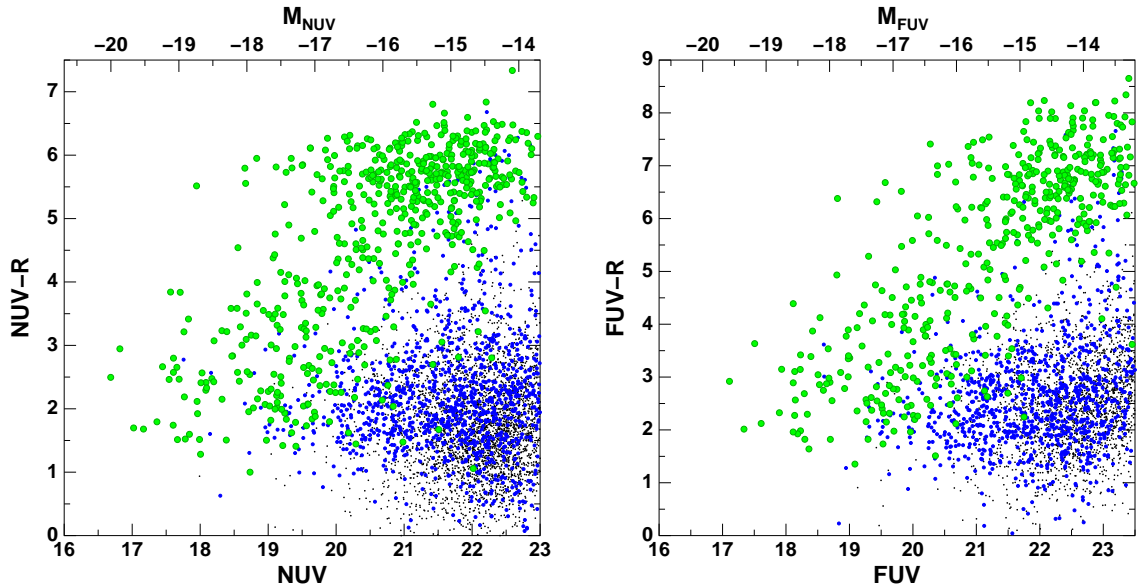}}
\caption{UV-optical colour-magntiude diagrams of galaxies in the Shapley supercluster core. Large green symbols indicate those galaxies spectroscopically confirmed as belonging to the supercluster (10\,$700<{\rm vel}<17$\,000\,km\,s$^{-1}$), blue symbols indicate those galaxies photometrically selected as supercluster members. Black dots indicate sources photometrically identified as background galaxies.}
\label{uv_cm}
\end{figure*}

To understand where and how this applies to our analyses, we show in Figure~\ref{uv_cm} the UV-optical colour-magnitude diagrams of galaxies in the Shapley supercluster core, in which the large green (small blue) symbols indicate galaxies with (without) redshift information placing them within the Shapley supercluster (10\,$700{<}cz{<}17$\,000\,km\,s$^{-1}$). 
In each panel we see the well known colour bimodality into the quiescent red sequence ($NUV{-}R{\simeq}6;\, FUV{-}R{\simeq}7$) and blue cloud  ($1{\la}{\rm NUV}{-}R{\la}4$) populations \citep[c.f.][]{haines08}. As might be expected, the brightest supercluster galaxies in the FUV and NUV all have colours indicative of actively star-forming galaxies.

In total 641 galaxies with both UV and $K$-band photometry have redshifts, of which 552 are supercluster members.
At bright magnitudes ($m_{AB}{<}18$), the estimation of the UV luminosity function of galaxies in the SSC can largely be performed spectroscopically, since to this point our redshift completeness is ${\ge}90$\%, falling to ${\sim}50$\% over $18{<}m_{AB}{<}20$. At fainter magnitudes, it is notable that while we have redshift information for the vast majority of the quiescent galaxy population ($NUV{-}R{>}4.5$) down to (and beyond) our UV completeness limits, for the star-forming ``blue cloud'' populations our redshift completeness drops rapidly beyond $NUV{\sim}19.5$ and $FUV{\sim}20$. This difference is due to the redshift surveys being optically-selected ($B_{J}$ or $R$ bands) and the substantial range in UV-optical colours between star-forming and quiescent populations. 
It will hence be necessary to quantify how many of the blue galaxies at faint UV magnitudes belong to the Shapley supercluster, and how many are background galaxies. Given that in each UV magnitude bin, the dominant contribution to the number counts will come from this blue star-forming population, this is the critical issue in determining the faint-end slope. 
Following \citet{cortese08}, we consider two different methods to estimate the UV galaxy luminosity function. 

\subsection{LF estimation via statistical subtraction of field galaxies}

In the first method we statistically account for field galaxies, per bin of UV magnitude, that are expected to lie within the area of interest by referring to the UV galaxy counts published by \citet{xu}.  By matching {\em GALEX} photometry from 35 MIS (Medium Imaging Survey) and 3 DIS (Deep Imaging Survey) pointings with optical star-galaxy classifications from the SDSS, they produced number counts of UV galaxies as a function of NUV and FUV magnitude over 24\,deg$^{2}$ to $m_{AB}{=}22.6$ and 1.7\,deg$^2$ over $22.6{<}m_{AB}{<}23.6$. We thus statistically account for contamination of background galaxies as a function of UV magnitude in bins of width 0.4\,mag over $14{<}m_{AB}{<}20$ and 0.2\,mag over $20{<}m_{AB}{<}23.6$ (matching the magnitude bins provided by Xu et al. 2005), and for each galaxy in a given bin estimate the probability that it belongs to the Shapley supercluster as $p_{i}=(N_{SC}-N_{f})/N_{SC}$, where $N_{SC}$ is the number of UV galaxies in the SSC survey region in that magnitude bin, and $N_{f}$ is the corresponding galaxy number count from \citet{xu} normalized to the Shapley UV survey area. Summing all of the $p_{i}$ within a given magnitude bin, we reobtain the statistical estimate for the number of supercluster galaxies in that bin, such that $p_{i}$ can be considered to be the contribution of each galaxy to the UV LF. For those galaxies with available redshifts, $p_{i}=1$ for 10\,$700{<}cz{<}17$\,000\,km\,s$^{-1}$ or 0 otherwise. 

\begin{figure*}
\centerline{\includegraphics[width=150mm]{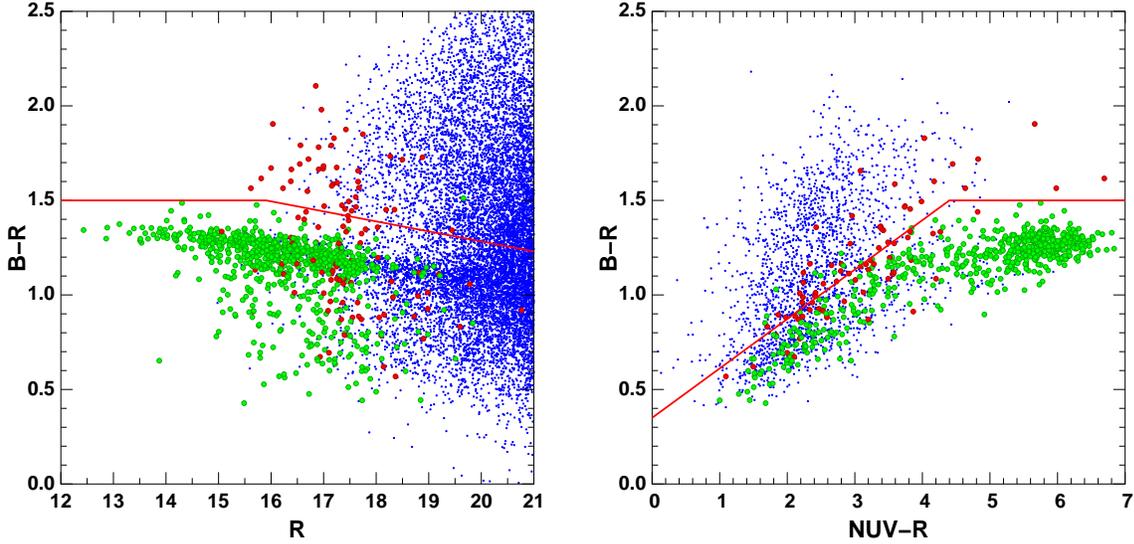}}
\caption{UV-optical colour selection criteria to identify supercluster members. {\em (left panel)} $NUV-R/B-R$ UV-optical colours. {\em (right panel)} $B-R/R$ colour magnitude diagram. Large green symbols indicate galaxies spectroscopically confirmed to belong to the supercluster (10\,$700<{\rm vel}<17$\,000\,km\,s$^{-1}$), while red symbols correspond to galaxies known to lie behind the supercluster (${\rm vel}>17$\,000\,km\,s$^{-1}$). Blue dots indicate galaxies ($R<20$ in the left panel) without redshifts. The red lines indicate the selection criteria used to identify supercluster members based on (i) their UV-optical colours (left panel) and (ii) their position with respect to the cluster red sequence (right panel)}  
\label{galcols}
\end{figure*}

We determine the UV k-corrections for each galaxy on the basis of its FUV, NUV and (when available) $B$ photometry. For star-forming galaxies with $FUV{-}NUV{\sim}0.5$ (or $F_{\lambda}{\propto}\lambda^{-1}$), these k-corrections are negligible ($<$0.05\,mag), but for quiescent galaxies with $FUV{-}NUV{\sim}1$--3, these can reach 0.2--0.3\,mag at $z{=}0.048$ \citep[see Fig. 2 of ][]{treyer}. Note that rather than fit model spectral energy distributions (SEDs) to the galaxy colours to estimate the UV k-correction, we simply fit power laws through the two adjacent bandpasses ($F_{\lambda}{\propto}\lambda^{\beta}$), which is generally a reasonable description of a star-forming galaxy's SED in the UV. We take the local spectral index for the FUV band (from which we derive the k-correction) from the spectral index corresponding to the FUV-NUV colour, and in the case of NUV take the local spectral index to be the average of that obtained from the FUV-NUV and NUV-B colours (if the latter is available). 

The primary source of uncertainty (and one which is often neglected) in estimates of cluster galaxy LFs is often the effect of clustering and cosmic variance on the background faint galaxy counts. This effect should be somewhat lower in the ultraviolet than in optical or infrared surveys of comparable depths, as ultraviolet-selected galaxies are less clustered than samples selected at longer wavelengths, and also tend to lie at lower redshifts. \citet{milliard} show that the median redshift of $NUV{<}22$ galaxies is 0.25, with little contribution from galaxies at $z{>}0.4$, hence any large-scale structures at high redshifts should have little or no effect on our ultraviolet galaxy counts. The contribution of cosmic variance to the uncertainty in galaxy counts within a survey region $\Omega$ can be estimated from the angular correlation function $w(\theta)$ as $\sigma^{2}=n\Omega + n^{2}\int d\Omega_{1} d\Omega_{2} w(\theta_{12})$, where $n$ is the number density of galaxies, and $d\Omega_{1}$ and $d\Omega_{2}$ are elements of the solid angle $\Omega$ separated by an angle $\theta_{12}$ \citep{ellis}. We include the effects of large-scale structure in the uncertainty in background galaxy counts in each magnitude bin using the the $NUV$ and $FUV$ angular correlation functions of \citet{milliard}.

The FUV (blue symbols) and NUV (red symbols) galaxy luminosity functions for the whole 2.8334\,deg$^{2}$ region covered by both {\em GALEX} and WFCAM $K$-band imaging are presented as open symbols in Fig.~\ref{uv_lf}, extending to $M_{\rm NUV}{=}{-}14$ and $M_{\rm FUV}{=}{-}13.5$. We fit single Schechter functions (shown as dashed curves) via a $\chi^{2}$-minimization routine, obtaining best-fit values $M^{*}_{\rm FUV}{=}{-}18.94^{+0.11}_{-0.18}$ and $\alpha_{FUV}{=}{-}1.67{\pm}0.03$ for the FUV LF, and $M^{*}_{\rm NUV}{=}{-}19.49^{+0.25}_{-0.31}$ and $\alpha_{\rm NUV}{=}{-}1.67{\pm}0.035$ for the NUV LF. The single Schechter fit provides a good description for both the FUV and NUV LFs down to $M_{\rm NUV}{=}{-}14$ and $M_{\rm FUV}{=}{-}13.5$, with $\chi^{2}_{\nu}{\sim}0.9$ for 10 dof.

\subsection{LF estimation via UV-optical colour selection}
\label{uvlfcolsel}

In the second method we use UV-optical colour selection criteria to exclude as many of the background galaxies as possible, and then correct for background contamination among those galaxies with colours consistent with the spectroscopically-confirmed supercluster population. 
\citet{cortese08} found that Coma cluster members could be efficiently distinguished from most background galaxies in the $g-i$ versus ${\rm NUV}-i$ observed colour-colour plot. In the left panel of Fig.~\ref{galcols} we show the $B-R$ versus ${\rm NUV}-R$ observed colour-colour plot used to produce an analogous separation between supercluster members and the background population, while in the right panel we show the $B-R/R$ colour-magnitude relation. The spectroscopically-confirmed supercluster members (large green symbols) lie within a well-defined region of the left-hand plot, while spectroscopically-confirmed background galaxies (${\rm vel}{>}17\,000$km\,s$^{-1}$; large red symbols) generally have redder $B-R$ colours than supercluster members with the same ${\rm NUV}-R$ colour, although there remains significant overlap. We thus identify by eye suitable selection criteria to produce a compromise between simultaneously minimize the number of supercluster members excluded and the number of background galaxies included as indicated by the red lines. We also apply a second criterion, excluding galaxies lie above the cluster red sequence (right panel), where among the galaxies with redshift information, all are found to be background galaxies.
Using this method we are able to exclude the bulk of the remaining galaxies for which we have no redshift information (blue points), although as apparent from the number of spectroscopically-confirmed background galaxies satisfying our selection criteria, there remain a significant level of background contamination among our colour-selected supercluster population. 

\begin{figure}
\centerline{\includegraphics[width=80mm]{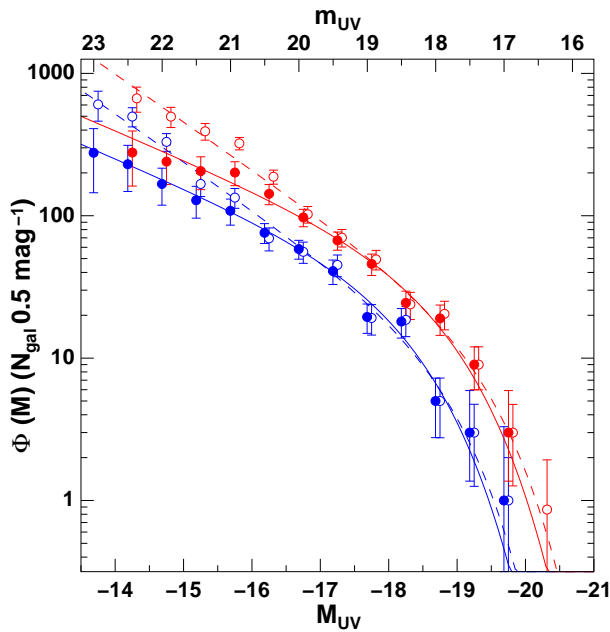}}
\caption{The FUV (blue) and NUV (red) LFs for the whole 2.8334\,deg$^2$ region covered by both {\em GALEX} and WFCAM $K$-band imaging obtained using: (i) the statistical subtraction of field galaxy method (open circles); and (ii) UV-optical colour selection (solid circles). The dashed and solid lines indicate the best-fitting Schechter functions to the data using the two methods.}
\label{uv_lf}
\end{figure}

Figure~\ref{galcols} shows that even with these colour cuts, a significant fraction of the remaining galaxies are background contaminants.
We attempt to account for these statistically, by measuring the real fraction, $f_{SC}(R)$, of supercluster members among those galaxies with redshifts satisfying our UV-optical colour selection criteria for each bin of unit $R$-band magnitude. This fraction declines slowly from 1 for $R{<}15$ to $0.80{\pm}0.04$ for $17{<}R{<}18$ and $0.65{\pm}0.10$ for $18{<}R{<}19$. Following the completeness-correction method of de Propris et al. (2003), we assume that within each $R$-band magnitude bin  the fraction of galaxies that are supercluster members is the same in the spectroscopic subsample as in the photometric one. This should be true if the spectroscopic targets were selected according only to their $R$-band magnitude and not their colour, which is largely true for each of the redshift surveys that contribute to our spectroscopic sample. It is important to note that we cannot apply a simple completeness correction to galaxies as a function of their UV magnitude, as the spectroscopic sample within a given UV magnitude bin is heavily biased towards the quiescent population with ${\rm NUV}-R>5$. 
 
\begin{figure}
\centerline{\includegraphics[width=70mm]{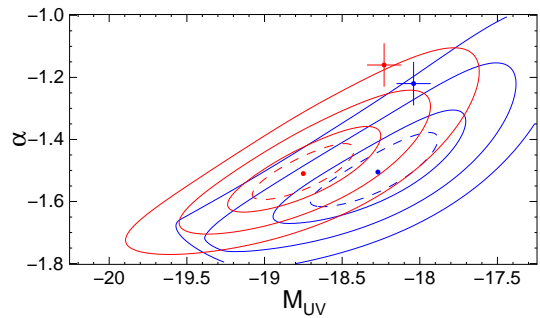}}
\caption{The confidence intervals of the Schechter function fits projected into the $M^{*}{-}\alpha$ plane for the FUV (blue) and NUV (red) LFs shown in Fig.~\ref{uv_lf}. The solid contours delineate $\chi^{2}-\chi^{2}_{min}=2.30$, 6.17 and 11.8, corresponding to the joint 1, 2 and 3$\sigma$ uncertainties on $M^{*}$ and $\alpha$, while the dashed contours indicate $\chi^{2}-\chi^{2}_{min}=1.0$. The corresponding local field galaxy FUV and NUV LFs of \citet{wyder} are also shown as blue and red points respectively with error bars.}
\label{uv_lffits}
\end{figure}

The FUV (blue) and NUV (red) galaxy luminosity functions obtained using this second method is shown by the solid symbols in Fig.~\ref{uv_lf}. The requirement for optical data to perform the UV-optical colour cuts means that these LFs are based on the 2.26\,deg$^{2}$ region covered by both the {\em GALEX} and SOS optical data, rather than the slightly larger region covered by WFCAM $K$-band imaging (see Fig.~\ref{map}). The error bars include Poisson uncertainties as well as uncertainties in $f_{SC}(R)$.  
We fit single Schechter functions (solid curves) via a $\chi^{2}$-minimization routine, obtaining best-fit values $M^{*}_{\rm FUV}{=}{-}18.27_{-0.39}^{+0.43}$ and $\alpha_{FUV}{=}{-}1.505{\pm}0.12$ for the FUV LF, and $M^{*}_{\rm NUV}{=}{-}18.76{\pm}0.31$ and $\alpha_{\rm NUV}{=}{-}1.505{\pm}0.085$ for the NUV LF. 
Again the single Schechter fit provides a good description for the NUV and FUV LFs down to the survey limit.
The best-fit parameters and their errors, calculated as the range of solutions within 1.0 of the minimum $\chi^{2}$ are presented in Table~1 and compared to previous determinations for the local field \citep{wyder,budavari}, the Coma cluster \citep{cortese08} and Abell 1367 \citep{cortese05}.
The errors in $\alpha$ and $M^{*}$ are highly correlated, and Fig.~\ref{uv_lffits} shows the joint 1, 2 and 3$\sigma$ error contours projected into the $M^{*}{-}\alpha$ plane (solid curves), while the dashed contours indicate $\chi^{2}{-}\chi^{2}_{min}{=}1.0$. 

\begin{table}
\begin{minipage}{8cm}
\caption{Best-fitting parameters for the NUV and FUV LFs}
\begin{tabular}{cccc} \hline
Band & Sample & \multicolumn{2}{c}{Schechter parameters} \\
& & $M^{*}$ & $\alpha$ \\ \hline
NUV & Shapley & $-18.76{\pm}0.31$ & $-1.505{\pm}0.085$ \\
NUV & Coma\footnote{Cortese et al. (2008)} & $-18.50{\pm}0.50$ & $-1.77_{-0.13}^{+0.16}$ \\
NUV & Abell 1367\footnote{Cortese et al. (2005)} & $-19.77{\pm}0.50$ & $-1.64{\pm}0.21$ \\
NUV & Local field\footnote{Wyder et al. (2005)} & $-18.23{\pm}0.11$ & $-1.16{\pm}0.07$ \\
NUV & $0.07{<}z{<}0.13$ field\footnote{Budav\'{a}ri et al. (2005)} & $-18.54{\pm}0.15$ & $-1.12{\pm}0.10$ \\ \hline
FUV & Shapley & $-18.27_{-0.39}^{+0.43}$ & $-1.505{\pm}0.120$ \\
FUV & Coma & $-18.20{\pm}0.80$ & $-1.61_{-0.25}^{+0.19}$ \\
FUV & Abell 1367 & $-19.86{\pm}0.50$ & $-1.56{\pm}0.19$ \\
FUV & Local field & $-18.04{\pm}0.11$ & $-1.22{\pm}0.07$ \\ 
FUV & $0.07{<}z{<}0.13$ field & $-17.97{\pm}0.14$ & $-1.10{\pm}0.12$ \\ \hline
NUV & Shapley SF & $-18.76_{-0.31}^{+0.34}$ & $-1.36_{-0.11}^{+0.13}$ \\
FUV & Shapley SF & $-18.12_{-0.43}^{+0.37}$ & $-1.38_{-0.135}^{+0.155}$ \\ \hline
\end{tabular}
\end{minipage}
\label{uvfits}
\end{table}

The ultraviolet LFs produced now are ${\sim}2\sigma$ shallower than those obtained by the first method. We note that these two methods for the background correction are subject to different {\em systematic} errors, which are not fully quantified in the error bars. The difference between these results is likely a reflection of these systematics, and suggests that either the statistical subtraction of field galaxies is underestimating the background galaxies or the colour selection is missing a fraction of real members. We note that a number of previous claims of steep faint-end slopes in clusters obtained via the statistical subtraction method have been shown to be inconsistent with shallower slopes obtained via deep spectroscopic follow-up surveys \citep[e.g.]{rines,penny}. In these cases the steep faint-end slopes appear due largely to the contribution of background galaxies redder than the C-M relation, and which are robustly removed by our UV-optical colours selection method. The statistical subtraction method is particularly suspectible to systematic errors due to photometric calibration offsets or corrections for Galactic extinction (which is large in our case). As field galaxy counts increase much more steeply than cluster member counts, small systematic uncertainties in background subtraction can produce large uncertainties in the numbers of faint cluster galaxies.
For both methods used, the NUV and FUV LFs of the Shapley supercluster galaxies are consistent within errors with those produced by \citet{cortese08} for the Coma cluster, while the $M^{*}_{UV}$ values for Abell 1367 appear a magnitude brighter, due to the presence in A\,1367 of a few galaxies with enhanced star formation, including an ultra-luminous UV galaxy.
These supercluster ultraviolet LFs are all significantly steeper than those obtained by \citet{wyder} for the local field galaxy population ($\alpha_{NUV}{=}{-}1.16{\pm}0.07$, $\alpha_{FUV}{=}{-}1.22{\pm}0.07$; red/blue points with error bars in Fig.~\ref{uv_lffits}). 

\begin{figure}
\centerline{\includegraphics[width=80mm]{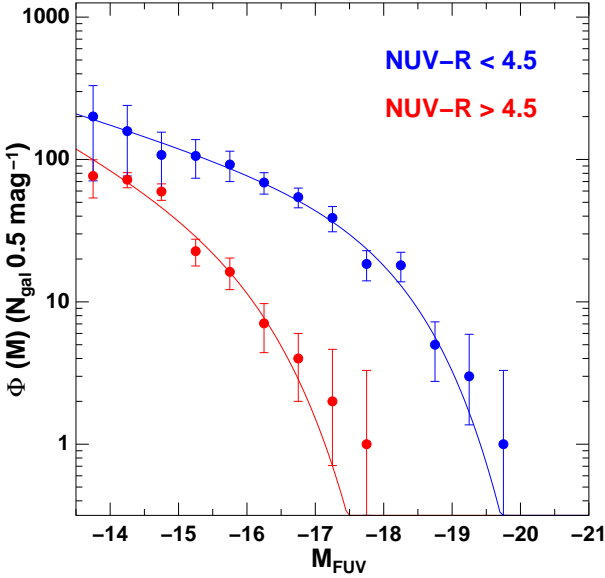}}
\caption{The FUV LFs for the passive ($NUV{-}R{>}4.5$; red symbols) and star-forming ($NUV{-}R{\le}4.5$; blue symbols) supercluster galaxy sub-populations. The solid lines indicate the best-fitting Schechter functions to the data.}
\label{fuv_lf_cols}
\end{figure}

Figure~\ref{fuv_lf_cols} examines the relative contributions of passive ($NUV{-}R{>}4.5$; red symbols) and star-forming ($NUV{-}R{\le}4.5$; blue symbols) galaxies to the far-UV luminosity function. 
We obtain best-fit Schechter functions for the star-forming supercluster galaxy populations of $M^{*}_{\rm FUV}{=}{-}18.12_{-0.43}^{+0.37}$, $\alpha_{FUV}{=}{-}1.38_{-0.135}^{+0.155}$ and $M^{*}_{\rm NUV}{=}{-}18.76_{-0.31}^{+0.34}$, $\alpha_{NUV}{=}{-}1.36_{-0.11}^{+0.13}$, consistent (within the 1$\sigma$ errors) with those obtained by \citet{wyder} and \citet{budavari} for local field galaxies. The contribution of the passive galaxies to the far-ultraviolet LF is negligible at the bright end ($M_{FUV}{\la}{-}17$), but becomes increasingly important at fainter magnitudes.

It is important to note that the FUV emission from these ``passive'' galaxies is unlikely to be due to low-levels of recent or ongoing star-formation. 
In a {\em HST}/ACS far-UV imaging survey \citet{salim10} also find that $NUV{-}r{\sim}4.5$ separates the early-type galaxy population into passive red sequence galaxies and those with evidence of extended residual star-formation in the form of rings or spiral arms. The FUV emission of these galaxies appears extended in the GALEX images, qualitatively ruling out a dominant AGN contribution.
\citet{haines08} show that there is very little contamination of the $NUV{-}r$ red sequence by galaxies spectroscopically classified as star-forming (with H$\alpha$ equivalent widths ${>}2${\AA}) in the local universe, while we find just one galaxy classified as star-forming among the ${\sim}100$ $NUV{-}R$ red sequence galaxies within the AAOmega spectroscopic sample of \citet{smith07}. Using the Rose Ca{\sc ii} index, \citet{smith09} also excluded frosting by small mass fractions (1--5\%) of recent star-formation (0.5--1\,Gyr old) for the vast majority of their quiescent galaxies in the AAOmega sample, which make up our $NUV{-}R$ red sequence. Instead the near-UV flux in these quiescent galaxies is dominated by hot main-sequence stars close to the turnoff \citep{dorman}, while the far-UV radiation is thought to be produced mainly by old (${\ga}10$\,Gyr) low-mass, helium burning stars in extreme horizonal-branch and subsequent phases of evolution \citep{oconnell}. 

\section{Mid/far-infrared luminosity functions}
\label{mirlf}

\begin{figure}
\centerline{\includegraphics[width=80mm]{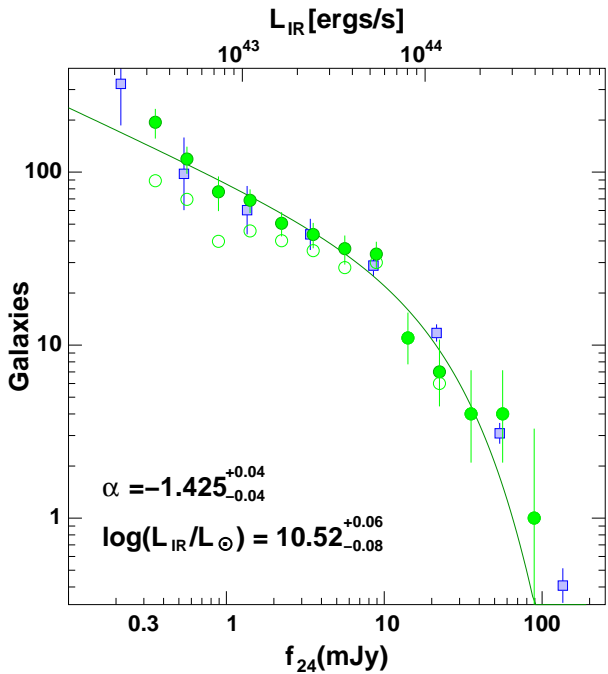}}
\caption{24$\mu$m luminosity function for the Shapley supercluster (solid green symbols). The contribution due to spectroscopically confirmed supercluster members is indicated by open symbols. Blue squares represent the field 24$\mu$m LF of \citet{marleau} for $0<z<0.25$ star-forming galaxies in the {\em Spitzer} First Look Survey.
Along the top axis we show the corresponding total infrared luminosity ($L_{IR}$[8--1000$\mu$m]), which for ease of comparison with the previous 24$\mu$m cluster luminosity functions of \citet{bai06,bai} we derive directly from the $L_{IR}{-}f_{24}$ relation used by \citet{bai} for the cluster A\,3266 at $z{=}0.06$, after correcting for the change in expected 24$\mu$m flux levels between Shapley and A\,3266. 
}
\label{lf24}
\end{figure}

We calculate the 24$\mu$m luminosity function for the 2.25\,deg$^{2}$ region covered by 24$\mu$m, optical and {\em GALEX} UV data. It is determined using the same completeness-correction method described in \S~\ref{uvlfcolsel},
identifying possible supercluster members from their UV-optical colours, and statistically accounting for background galaxies using the existing spectroscopic information. To account for the incompleteness of the 24$\mu$m data, we use the inverse of the completeness function (shown in Figure~\ref{compl_24um}) as weighting factors when we calculate the number counts in each bin in 24$\mu$m flux. 
The resulting 24$\mu$m luminosity function, shown by the solid green points in Fig.~\ref{lf24}, covers a factor 300 in luminosity, extending down to our survey limit of 0.35\,mJy. The open symbols indicate the contribution just from the 360 galaxies with $f_{24}{>}0.35$\,mJy and redshifts placing them in the Shapley supercluster, which represents ${>}80$\% of galaxies with $f_{24}{>}8$\,mJy, falling to ${\sim}4$0\% at the completeness limit. We have redshifts for all $f_{24}{>}26$\,mJy galaxies, and so are not missing any LIRGs in the supercluster, irrespective of their UV--optical colours.

To estimate the total infrared luminosity, $L_{IR}$(8--1000$\mu$m), from our 24$\mu$m fluxes, we adopt the calibration used by \citet{bai} for the cluster A\,3266 at $z{=}0.06$ (after accounting for the small redshift difference between the Shapley supercluster and A\,3266), which was derived by comparison to the sample of star forming galaxy SEDs developed by \citet{dh02}. These SEDs are based on {\em IRAS} and {\em Infrared Space Observatory} observations of 69 normal star-forming galaxies \citep{dale} and further calibrated with submillimeter observations. The SEDs are luminosity dependent, such that the 24$\mu$m contribution becomes increasingly important with $L_{IR}$, resulting in the linear relation $\log L_{IR}({\rm erg\,s}^{-1}){=}42.91+0.86{\times}\log f_{24}$(mJy) at the distance of Shapley (shown along top axis of Fig.~\ref{lf24}). 

We obtain a best-fit Schechter function of $f_{24\mu{\rm m}}^{*}{=}24.83^{+3.35}_{-4.65}$mJy and $\alpha(24\mu{\rm m}){=}{-}1.425^{+0.035}_{-0.040}$, shown by the dark green curve in Fig.~\ref{lf24}. Adopting the $L_{IR}{-}f_{24}$ relation derived above, these correspond to $\log (L_{IR}^{*}/L_{\odot}){=}10.52^{+0.06}_{-0.08}$ and $\alpha(IR){=}{-}1.49{\pm}0.04$. 
These values are fully consistent with both those obtained by \citet{bai06} for Coma ($\log (L_{IR}^{*}/L_{\odot}){=}10.48^{+0.48}_{-0.31}$ and $\alpha(IR){=}{-}1.49{\pm}0.11$) and \citet{bai} for Abell 3266 ($\log (L_{IR}^{*}/L_{\odot}){=}10.49^{+0.13}_{-0.11}$ for a fixed $\alpha=-1.41$). 

We now compare our supercluster 24$\mu$m galaxy LF with that for local field galaxies by overlaying as blue squares the 24$\mu$m luminosity function obtained by \citet{marleau} for $0{<}z{<}0.25$ star-forming galaxies in the {\em Spitzer} First Look Survey. We find the shape of the 24$\mu$m LF of Shapley supercluster galaxies to be fully consistent with that obtained for local field galaxies, and support the assertion of \citet{bai} that there is no environmental dependence on the shape of the 24$\mu$m luminosity function. \citet{marleau} find evidence for an upturn in the galaxy counts below $\nu L_{\nu}^{24\mu m}{\sim}10^{8}L_{\odot}$ ($L_{IR}{\sim}5{\times}10^{42}$\,erg\,s$^{-1}$), particularly once they include the contribution of galaxies with only photometric redshift estimates, and we find marginal evidence for this upturn, albeit only in the faintest 1--2 luminosity bins, where our spectroscopic completeness is lowest and photometric uncertainties are greatest. \citet{westra} also find a similar upturn in the local ($0.01{<}z{<}0.10$) extinction-corrected H$\alpha$ luminosity function, below $L(H\alpha){\sim}10^{40}$\,erg\,s$^{-1}$ (${\rm SFR}{\sim}0.1{\rm M}_{\odot}\,{\rm yr}^{-1}$). 

\begin{figure}
\centerline{\includegraphics[width=80mm]{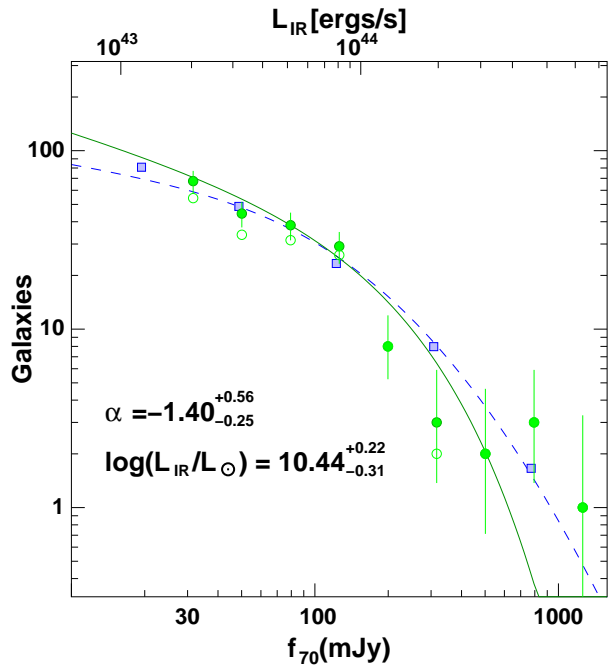}}
\caption{70$\mu$m luminosity function for the Shapley supercluster (solid green symbols). The contribution due to spectroscopically confirmed supercluster members is indicated by open symbols. The best fitting Schechter function is indicated by the solid green curve and the associated parameters and uncertainties indicated. The blue squares indicate the local field luminosity function of IRAS galaxies from \citet{wang}, while the blue dashed curve indicates the analytic form of the local field IR luminosity function of \citet{takeuchi}, both obtained from the PSCz survey.}
\label{lf70}
\end{figure}

We do not present an estimate of the 24$\mu$m supercluster LF using the background subtraction method, even though robust measures of 24$\mu$m galaxy counts down to our survey limit exist. \citet{shupe} presented 24$\mu$m galaxy counts from the six SWIRE fields, covering in total 64\,deg$^{2}$, but showed that even with these large fields (5--11\,deg$^{2}$) containing tens of thousands of galaxies, the field-to-field differences were of the order 10\%, much larger than could be accountable to Poisson uncertainties. In the sub-mJy range (0.3--1\,mJy), \citet{shupe} find that only 10--20\% of 24$\mu$m extragalactic sources are at $z{<}0.3$, and so the field 24$\mu$m count variations, significant even on these 3 degree scales, may be due to large-scale structures at $z{\ga}0.5$. 
For our faintest 24$\mu$m luminosity bin, the field-to-field variance seen in the SWIRE galaxy counts (D. Shupe, private communication) corresponds to an uncertainty of ${\sim}400$, double the estimated cluster galaxy counts from Fig.~\ref{lf24}. 
The huge uncertainties due to cosmic variance can be understood as a consequence of just ${\sim}5$\% of the 24$\mu$m sub-mJy extragalactic sources in the Shapley field actually  belonging to the supercluster. Note that the uncertainies obtained using the completeness-correction method as indicated by the error bars in Fig.~\ref{lf24} are much smaller, due to a combination of the extensive spectroscopic information, and the ability of the UV-optical colour selection to exclude the bulk of the background 24$\mu$m sources. 

In Fig.~\ref{lf70} we present the corresponding 70$\mu$m luminosity function of Shapley supercluster galaxies (solid green symbols) using again the completeness-correction method of \S~\ref{uvlfcolsel}. As before, the open symbols indicate the contribution spectroscopically-confirmed supercluster members. Along the top axis we show the corresponding bolometric infrared luminosites based on the SEDs of \citet{rieke09}. We fit the 70$\mu$m luminosity function by a single Schechter function as shown by the green curve, obtaining best-fit parameters of $f_{70\mu{\rm m}}^{*}{=}193^{+127}_{-99}$mJy and $\alpha(70\mu{\rm m}){=}{-}1.40^{+0.56}_{-0.25}$. Using the \citet{rieke09} SEDs, we convert this to a bolometric infrared luminosity, obtaining $\log (L_{IR}^{*}/L_{\odot}){=}10.44^{+0.22}_{-0.31}$. This is the first measurement of the 70$\mu$m luminosity function of cluster galaxies with {\em Spitzer}, and appears fully consistent with that obtained at 24$\mu$m, albeit with large uncertainties in the Schechter parameters due to the relatively shallow depth. We compare the supercluster 70$\mu$m with the 60$\mu$m local field galaxy luminosity function obtained by \citet{wang} from the all-sky {\em IRAS} Point Source Redshift Survey (blue squares), assuming a MIPS-70$\mu$m/IRAS-60$\mu$m flux ratio of  1.56 based on the \citet{dale} infrared SED with $\alpha{=}2.5$, redshifted to $z=0.048$. This model SED best describes the $f_{70}/f_{24}$ mid-infrared colours of the majority of our 70$\mu$m-bright supercluster population (Paper III).
 The blue dashed line indicates the best-fit analytic function for the local 60$\mu$m LF obtained by \citet{takeuchi}. Again, we find no significant difference between the 70$\mu$m luminosity function of Shapley supercluster galaxies and that obtained for local field galaxies.

\section{Comparison of the UV and MIR emission from cluster galaxies}
\label{sec:uvir}

The infrared-to-ultraviolet ratio is a rough measure of the amount of extinction at ultraviolet wavelengths of young stellar populations (${\la}10^{8}$\,yr), with the caveat that the {\em observed} ultraviolet and infrared emission do not come from exactly the same regions within a star-forming region or galaxy, or be produced by recently formed stars of the same ages/masses \citep{calzetti05}. The main problem affecting UV and optical-based SFR indicators is dust obscuration, which can require corrections of the order 10--100 in the SFR estimator, particularly in the ultraviolet \citep{bell}. This correction is also sensitive to the metal content, star-formation history and the relative distribution and geometry of the interstellar dust with respect to the stars producing the ultraviolet emission \citep{calzetti05,cortese08,boquien}. 

\begin{figure}
\centerline{\includegraphics[width=80mm]{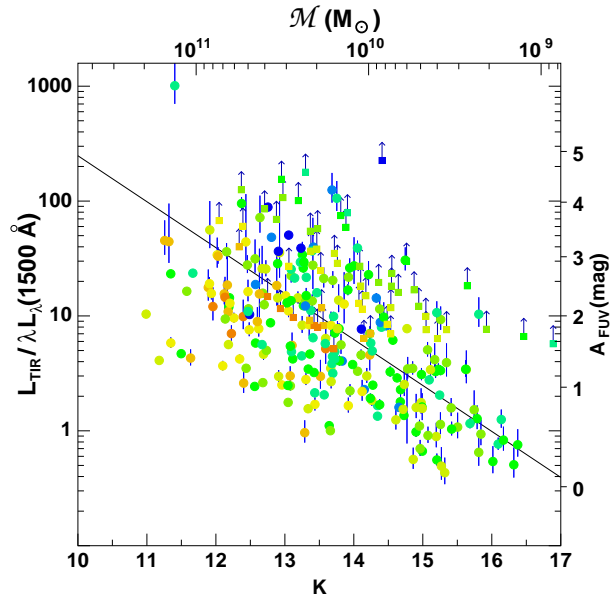}}
\caption{Infrared-to-ultraviolet ratio as a function of $K$-band magnitude for star-forming galaxies ($f_{24}/f_{K}{>}0.15$) in the Shapley supercluster. Symbol are coloured according to their $f_{24}/f_{K}$ ratio from yellow/orange (${<}0.8$) to blue (${>}8$). Lower limits are shown for galaxies fainter than our nominal far-ultraviolet completeness limit of $m_{FUV}{=}22.5$. The diagonal line indicates a trend of the form $L_{TIR}/L_{FUV}{\propto}L_{K}$. The far-ultraviolet extinction prescription used for the right-hand axis comes from \citet{buat}.}
\label{uvirtrends}
\end{figure}

In Figure~\ref{uvirtrends} we show the infrared-to-ultraviolet ratio as a function of $K$-band magnitude for those galaxies spectroscopically-confirmed as Shapley supercluster members, estimating the total infrared luminosity ($L_{TIR}$) for each galaxy from their 24$\mu$m fluxes using the infrared SED models of \citet{rieke09}. The right-hand axis shows the resultant level of attenuation in the FUV estimated from the infrared-to-ultraviolet ratio based on the prescription described in Eq.~2 from \citet{buat}. This is valid for star-formation histories in which the bulk of dust emission is related to the ultraviolet absorption, as for most normal star-forming galaxies, and any range of configuration of dust/star geometry, metallicity or attenuation law should affect the relation by less than 20 per cent \citep{buat}. In cases where a signifcant fraction of the dust heating comes from evolved stars, such as early-type galaxies, the level of attenuation in the FUV should be significantly lower for the same infrared-to-ultraviolet ratio \citet{cortese08b}. 
The colours of each symbol indicate the $f_{24}/f_{K}$ flux ratio, which can be considered a proxy for the specific-SFR of the galaxy, from orange/yellow (0.15--1), through green (1--8; lying along the star-forming sequence, consistent with being normal spiral galaxies) to blue (${>}8$; above the main star-forming sequence). In paper III, we give a full discussion of how the observed $f_{24}/f_{K}$ flux ratio relates to the specific-SFR and the bimodality between star-forming and passive galaxies.
In the case of passively-evolving galaxies neither the FUV nor the 24$\mu$m emission is believed to be primarily produced by star-formation, but rather evolved stars \citep[see Paper III;][]{oconnell,bressan}, and so we show only galaxies identified as star-forming according to having $f_{24}/f_{K}{>}0.15$.

The infrared-to-ultraviolet ratio correlates strongly with the $K$-band luminosity, increasing from ${\sim}1$ at $K{\sim}16$ ($\mathcal{M}{\sim}10^{9}\,{\rm M}_{\odot}$) to ${\sim}100$ for the most massive galaxies, consistent with an overall trend of the form $L_{TIR}/L_{FUV}{\propto}L_{K}$ (diagonal line). The trend for more massive galaxies to be more heavily attenuated in the ultraviolet than low-mass systems was previously described by both \citet{cortese06} and \citet{dale07}, although both these authors obtain somewhat shallower trends than found here.   
However, at fixed $K$-band luminosity, there remains a significant scatter, which also increases with $K$-band luminosity. These trends are consistent also with those obtained for A(H$\alpha$) by \citet{brinchmann} and \citet{garn} for starforming galaxies in the SDSS survey, as estimated from the Balmer decrement (H$\alpha$/H$\beta$). The primary driver behind the increase in extinction with stellar mass is thought to be a combination of the greater dust content of more massive galaxies (as would be produced assuming a constant dust gas fraction for example) and the mass-metallicity relation, such that more massive (and hence metal-rich) galaxies have higher dust-to-gas ratios \citep{munoz}.
Finally, at fixed $K$-band luminosity, the infrared-to-ultraviolet ratio increases with $f_{24}/f_{K}$, i.e. specific-SFR, the blue symbols appearing located systematically above the orange/yellow ones. 

\subsection{The global contributions of obscured and unobscured star-formation}

\begin{figure}
\centerline{\includegraphics[width=80mm]{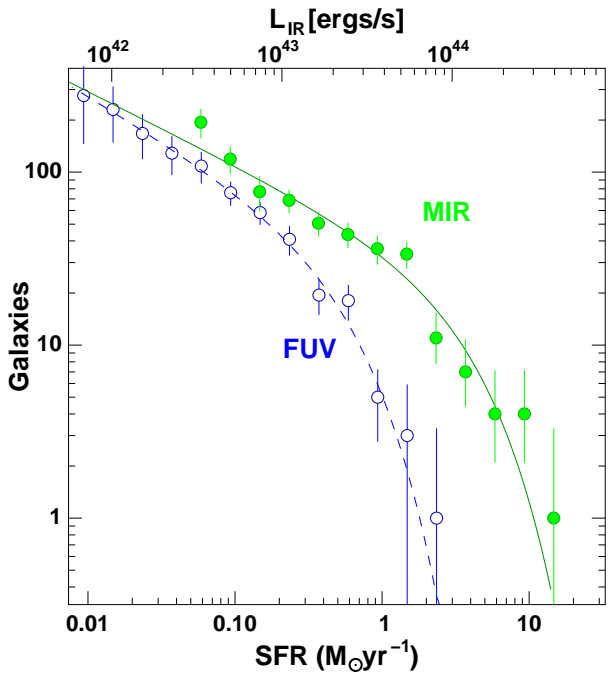}}
\caption{Comparison of the FUV and 24$\mu$m luminosity functions in terms of the derived SFRs, assuming the UV and MIR components of the SFR calibration of \citet{leroy}.}
\label{sfrlf}
\end{figure} 

As we have just seen, the star formation in many of the supercluster galaxies is heavily obscured, with more of the energy from young stars output in the infrared than the ultraviolet. How does this average out over the entire supercluster population? Using the ultraviolet and 24$\mu$m-based calibrations of \citet{leroy}
\begin{align}
{\rm SFR (M}_{\odot}{\rm yr}^{-1}) & = & 0.68{\times}10^{-28}L_{\nu}{\rm (FUV)[erg\,s}^{-1}{\rm Hz}^{-1}]\notag \\
& + & 2.14_{-0.49}^{+0.82}{\times}10^{-43}L(24\mu {\rm m)[erg\,s}^{-1}],
\label{leroy}
\end{align}
(which assumes a Kroupa IMF), 
 we can directly compare the observed FUV and 24$\mu$m luminosity functions in terms of star-formation rates, the result of which is shown in Figure~\ref{sfrlf}. 
We can immediately see that for the galaxies with the highest SFRs, the estimates obtained via the mid-infrared are ${\sim}10{\times}$ higher than those obtained from the uncorrected far-ultraviolet, and that for ${\rm SFRs}{\ga}1\,{\rm M}_{\odot}\,{\rm yr}^{-1}$, the far-UV component can be neglected and the 24$\mu$m LF can be used as a good estimator of the SFR distribution. At lower SFRs however (${\la}1\,{\rm M}_{\odot}\,{\rm yr}^{-1}$), the FUV component cannot be neglected, as it becomes equally important as the infrared component, comprising anywhere between 10--90 per cent of the emission from star-formation in these galaxies. This trend for the dust attenuation to increase with SFR is consistent with that observed by \citet{buat07} for local field galaxies, and suggests that for surveys limited to star formation rates ${\ga}1\,{\rm M}_{\odot}\,{\rm yr}^{-1}$, mid-infrared observations alone should be sufficient, encompassing ${\ga}90$ per cent of the emission from ongoing star-formation.

Over the entire SOS region, we estimate a total supercluster SFR of $327^{+102}_{-60}\,{\rm M}_{\odot}\,{\rm yr}^{-1}$, of which $264^{+102}_{-60}\,{\rm M}_{\odot}\,{\rm yr}^{-1}$ (${\sim}80$ per cent) is obscured (based on the 24$\mu$m calibration) and just $63{\pm}3\,{\rm M}_{\odot}\,{\rm yr}^{-1}$ (${\sim}20$ per cent) is unobscured and emitted in the form of ultraviolet continuum. For the mid-infrared contribution we excluded those galaxies with $f_{24}/f_{K}{<}0.15$ for which we assume the 24$\mu$m emission comes from evolved stars given that all such galaxies within the spectroscopic sample of \citet{smith07} are classified as passive based on their lack of H$\alpha$ emission (Paper III), and have colours similar to early-type galaxies whose infrared emission has been shown to come from the dusty circumstellar envelopes of mass-losing AGB stars \citep{bressan,clemens}. 

The infrared error is entirely dominated by the 30\,per cent uncertainty in the SFR calibration of \citet{leroy}, which is comparable to the differences among the most widely used 24$\mu$m SFR calibrations in the literature \citep[for a comparison of these see e.g.][]{calzetti10}. We note that the 24$\mu$m calibration of \citet{leroy} is consistent to that of \citet{rieke09} which is SFR(M$_{\odot}\,{\rm yr}^{-1}){=}2.04{\times}10^{-43}L(24{\mu}{\rm m})[{\rm erg s}^{-1}]$, as well as those of \citet{wu} and \citet{zhu}, all of which are based on global galaxy measurements. Using instead the SEDs of \citet{rieke09} to estimate the total infrared luminosity from the 24$\mu$m fluxes and the infrared SFR calibration of \citet{buat08}, we obtain SFRs that are systematically 10--20\,per cent lower, once we account for the expected 30\,per cent contribution to the infrared flux from evolved stars \citep{bell03,iglesias}, although we note that this relative contribution is likely to depend on the particular star-formation histories of the galaxies involved \citep[e.g.][]{cortese08b,calzetti10}. Even better consistency is found if we use instead the $L_{TIR}$-SFR calibration of \citet{bell03}, producing SFRs within 10\,per cent of those based on Eq.~\ref{leroy} for ${\sim}10^{10}-10^{11}\,L_{\odot}$ after accounting for the different IMF used. 

The remaining sources of uncertainty contribute to an uncertainty in the SFR$_{IR}$ of just 11\,M$_{\odot}\,{\rm yr}^{-1}$. These include the 4\% absolute calibration uncertainty of MIPS at 24$\mu$m \citep{engelbracht}, the statistical uncertainty arising from including the photometrically-selected supercluster population, and the effects of varying the cut-off in the range $0.1{\la}f_{24}/f_{K}{<}0.2$. For the far-ultraviolet contribution we excluded those galaxies with $NUV-R{>}4.5$ (the emission again due to evolved stars; see {\S}\,3.2), and the error estimate includes the 0.05\,mag uncertainty in the $FUV$ absolute calibration \citep{morrissey07} and statistical uncertainty from the photometrically-selected supercluster members. 

We believe the bulk of the mid-infrared emission (and hence our SFR budget) to be due to star formation, rather than AGN, whose contribution we estimate as of the order 5--10 per cent. This we base on the observed tight FIR--radio correlation seen for the far-IR bright cluster members (Paper III) when comparing both the 24$\mu$m and 70$\mu$m fluxes with the published 1.4\,GHz catalogue of \citet{miller}. Moreover, these galaxies have FIR--radio colours more consistent with dust heated by low-intensity star formation rather than AGN, and have extended 24$\mu$m emission. However in Paper III, we find four 24$\mu$m-bright sources whose emission remains unresolved by MIPS, and whose FIR--radio colours are more consistent with dust heated by AGN. Two of these are within the spectroscopic sample of \citet{smith07} and are classified as AGN, and between then contribute ${\sim}15\,{\rm M}_{\odot}\,{\rm yr}^{-1}$ to the cluster SFR budget. Unfortunately, we do have {\em Spitzer}/IRAC photometry, to identify IR-bright AGN from their power-law spectra in the 3--10$\mu$m range \citep{stern}, making a complete census of the AGN population impossible. There is however partial X-ray coverage of the SCC via a mosaic of 10 XMM images, each with exposure times of 20--40\,ksec, covering A\,3562, A\,3558, SC\,1329-313 and SC\,1327-312 and most of the connecting filamentary strucutre, allowing us to identify AGN as X-ray point sources. Based on a sample of 202 spectroscopic SSC members detected at 24$\mu$m, comprising 52\,per cent of the 24$\mu$m flux associated to SSC galaxies, lying within the XMM mosaic, we estimate that $10.6^{+5.2}_{-4.2}$\,per cent of the 24$\mu$m flux comes from X-ray loud AGN. We note that in infrared-bright sources, often both AGN and starburst activity are ongoing concurrently, and that the contribution from the AGN component (typically 5--50 per cent) can only be resolved via infrared spectroscopy \citep[e.g.][]{goulding}, and so we believe that this estimate of ${\sim}1$0\,per cent AGN contamination represents an upper limit.

\section{Discussion}\label{sec:discuss}

The luminosity function is one of the most basic and fundamental properties of the galaxy population, and its form is shaped by many of the processes of galaxy formation \citep{benson}. The optical and near-infrared luminosity functions of field and cluster galaxies are well described by the \citet{schechter} function, which has a sound theoretical basis as the expected mass function of galaxies built up by gravitational hierarchical assembly, although it seems that additional feedback processes from AGN and supernovae are required to fit the observed sharp cut off at bright luminosities and the shallow faint-end slope \citep{bower}. As \citet{yun} argue, if star formation activity is an integral part of galaxy evolution and reflects the nature of the host galaxies, then the Schechter function should also offer as good a description for the ultra-violet, far-infrared and radio LFs as it does for the optical and near-infrared LFs.

In Sections~\ref{uvlf} and~\ref{mirlf} we have obtained the ultraviolet and far-infrared galaxy luminosity functions for the Shapley supercluster, complementing the existing optical \citep{sos1} and near-infrared \citep{merluzzi} LFs for the same region. In the ultraviolet, using new {\em GALEX} imaging, we find that the NUV ($\lambda_{e}{=}2310${\AA}) and FUV ($\lambda_{e}{=}1510${\AA}) LFs are both well described by single Schechter functions down to $M_{UV}{=}{-}14$ (${\sim}M^{*}{+}4.5$) with $M_{FUV}{=}{-}18.27{\pm}0.41$, $\alpha_{FUV}{=}{-}1.505{\pm}0.12$ for the far-UV, and $M_{NUV}{=}{-}18.76{\pm}0.31$, $\alpha_{NUV}{=}{-}1.505{\pm}0.085$ for the near-ultraviolet. These LFs are essentially identical once the typical ultraviolet colours of star-forming galaxies found in the supercluster having $FUV{-}NUV{\sim}0.5$ ($F_{\lambda}{\propto}\lambda^{-1}$) are taken into account. These ultraviolet LFs of the Shapley supercluster core are fully consistent with those obtained by \citet{cortese03,cortese05,cortese08} for Virgo, Coma and Abell 1367. 

The faint-end slopes of local cluster ultraviolet LFs ($\alpha{\sim}{-}1.5$) are all observed to be significantly steeper (at the ${\sim}2{\sigma}$ level) than the local field galaxy UV LFs of \citet{wyder} and \citet{budavari}, implying that the shape of the LF depends on environment (cluster versus field). This appears at least partly due to the morphology-density or star formation-density relations, as Fig.~\ref{fuv_lf_cols} shows that the blue, star-forming ($NUV{-}R{<}4.5$) and quiescent ($NUV{-}R{>}4.5$) galaxies contribute to the LFs in quite different ways. The blue, star-forming galaxies dominate at the bright-end, and have a faint-end slope $\alpha_{FUV}{=}1.38_{-0.135}^{+0.155}$ consistent within errors with the local field galaxy FUV LF (albeit still marginally steeper).  
The contribution of the passive galaxies to the far-ultraviolet LF is negligible at the bright end ($M_{FUV}{\la}{-}17$), but becomes increasingly important at fainter magnitudes.  This pile-up of quiescent galaxies in the last three magnitude bins ($M_{FUV}{>}{-}15$) appears the primary cause of the discrepancy in the faint-end slopes between cluster and field galaxies. This effect is seen also in the Coma cluster ultraviolet LF \citep{cortese08},  and also at optical wavelengths \citep[e.g.][]{sos1}. 
A second contributing factor to the steeper observered cluster UV LFs could be simply the steeper stellar mass functions (or $K$-band LFs) of cluster galaxies, as observed by \citet{merluzzi} for this same Shapley supercluster core region, finding $\alpha_{K}{=}{-}1.42{\pm}0.03$ as opposed to $\alpha_{K}{=}{-}1.16{\pm}0.04$ for the local field $K$-band LF \citep{jones}. Given that the FUV emission from the ``passive'' galaxies is unlikely to be due to star-formation \citep{oconnell,dorman}, the relative consistency between the FUV LFs of cluster star-forming galaxies and the field, suggests little environmental dependance in the unobscured SFRs of star-forming galaxies.   

In {\S}~\ref{mirlf} using new panoramic {\em Spitzer}/MIPS 24$\mu$m imaging, we determined the infrared galaxy LF for the Shapley supercluster core (Fig.~\ref{lf24}), finding it well described by a single Schechter function with $\log (L_{IR}^{*}/L_{\odot}){=}10.52^{+0.06}_{-0.08}$ and $\alpha(IR){=}{-}1.49{\pm}0.04$. These values are fully consistent with both those obtained by \citet{bai06} for Coma ($\log (L_{IR}^{*}/L_{\odot}){=}10.48^{+0.48}_{-0.31}$ and $\alpha(IR){=}{-}1.49{\pm}0.11$) and \citet{bai} for Abell 3266 ($\log (L_{IR}^{*}/L_{\odot}){=}10.49^{+0.13}_{-0.11}$ for a fixed $\alpha=-1.41$). Although we do not extend to quite as low luminosities as \citet{bai} our larger sample size allows us to derive stronger constraints on the 24$\mu$m luminosity function of local cluster galaxies. 

The {\em Spitzer}/MIPS 24$\mu$m band is some way from the typical peak wavelength (${\sim}100{\mu}$m) of the far-infrared emission from galaxies, requiring uncertain extrapolations in estimating $L_{IR}$ from the 24$\mu$m flux such as via the model infrared SEDs of \citet{dh02} or \citet{rieke09}. The {\em Spitzer}/MIPS 70$\mu$m filter, being closer to the far-infrared peak affords more reliable $L_{IR}$ estimates, and so we have redetermined the infrared LF, based on our available 70$\mu$m photometry of the Shapley supercluster core (Fig.~\ref{lf70}). A best-fit Schechter function with $\log (L_{IR}^{*}/L_{\odot}){=}10.44^{+0.22}_{-0.31}$ and $\alpha(IR){=}{-}1.40_{-0.25}^{+0.56}$ is obtained, fully consistent with the 24$\mu$m LF, albeit with large uncertainties due to the relatively shallow depth of our 70$\mu$m imaging. This represents the first measurement of the 70$\mu$m galaxy LF of a local cluster with {\em Spitzer}, and should prove a useful local benchmark to follow the evolution in the infrared properties of cluster galaxies with ongoing and future surveys \citep[e.g.][]{haines09,haines09b,haines10,smith10,braglia,rawle,chung} with {\em Spitzer} and particularly {\em Herschel} whose PACS instrument covers rest-frame 70$\mu$m beyond $z{\sim}1$.

\citet{bai06} found that the 24$\mu$m galaxy LF of Coma was consistent with the infrared LFs of local field galaxies, based on comparison to the 12$\mu$m LF derived from the all-sky {\em IRAS} Faint Source Catalogue \citep{rush} or the 60$\mu$m LF of local galaxies in the {\em IRAS} Point Source Redshift Survey \citep{takeuchi}. We also find the shape of both the 24$\mu$m and 70$\mu$m LFs of Shapley supercluster galaxies to be fully consistent with that obtained for local field galaxies by \citet{marleau} at 24$\mu$m (blue squares in Fig.~\ref{lf24}) and \citet{takeuchi} and \citet{wang} at 70$\mu$m (dashed line and blue squares respectively in Fig.~\ref{lf70}), confirming the assertion of \citet{bai} that there is no environmental dependence on the shape of the far-infrared luminosity function. 

A probable significant consequence of this apparent complete agreement between the local cluster and field galaxy infrared LFs, is that the bulk of star-forming galaxies that make up the observed cluster infrared LFs have been recently accreted from the field and haven't had their star formation activity significantly affected by the cluster environment yet, based on a similar argument to that used by \citet{balogh04b} regarding the lack of environmental dependence seen for the distribution in EW(H$\alpha$). If the mechanism which quenches star-formation in infalling galaxies acts rapidly, then these galaxies rapidly drop out of the sample, and no longer contribute to the far-infrared LF (except perhaps at the very lowest luminosities studied here). The primary change to the infrared LF is then a reduction in the overall normalization, but without affecting its shape (i.e. $L^{*}$ or $\alpha$). 
If instead, the quenching of star-formation acts slowly (${\ga}1$\,Gyr), then a significant fraction of the cluster galaxies which constitute the infrared LF will be those {\em in the process of} being quenched, and have reduced (often by a factor 2 or more) SFRs and infrared luminosities in comparison to the field population. The primary consequence of there existing a significant population of transforming galaxies would be that the faint-end slope {\em steepens}, while the Schechter function becomes an increasingly poor fit to the data. However, the shape of the infrared LF is not expected to be a particularly sensitive indicator to the presence of galaxies being quenched by environmental processes, based on similar arguments to those of \citet{cortese08} about the ultraviolet luminosity function.

The bulk of these infalling star-forming galaxies cannot be interlopers, who happen to have line-of-sight velocities within the redshift range of the cluster,   as the normalization of the cluster infrared LF is still ${\sim}2$5--3$0{\times}$ higher than the local field infrared LFs \citep{marleau,wang} multiplied by the total volume occupied by the cluster survey (${\sim}2$\,400\,Mpc$^{3}$). We note that this higher normalization does not imply star formation is being triggered in the supercluster, but simply reflects the overall significant overdensity of galaxies in this volume. 

It is difficult to understand why if the field and cluster infrared LFs are consistent, the cluster FUV LF should be significantly steeper than that found in the field, as both reflect the star-formation from the essentially same galaxies. One possibility is that the differences are only fully apparent at the very low SFRs traced by the GALEX data, and which are missed by the marginally ($\sim$2--3$\times$) shallower 24$\mu$m data. Alternatively, the skewed FUV LF could reflect environmental processes acting on just dwarf galaxies which are missed by our 24$\mu$m survey, as their star formation is unobscured and emitted primarily in the FUV.

One small caveat to the apparent agreement between cluster and field infrared LFs over the luminosity range $10^{9}-10^{11}L_{\odot}$ is the observed behaviour of the local field far-infrared LF, which for $L_{IR}{\ga}10^{11}L_{\odot}$ can no longer be described by a single Schecter function. 
Instead of declining exponentially, the luminosity function is better described as a power-law \citep{yun,takeuchi}, or via an additional Schechter function to model the galaxy counts at $L_{IR}{\ga}10^{11}L_{\odot}$. Similarly, the 1.4\,GHz radio galaxy luminosity function cannot be described by a single Schechter function, but requires two components, one representing the normal, late-type field galaxies and the other representing ``monsters'' powered by AGN \citep{condon,mauch} which dominate at $L_{1.4\,{\rm GHz}}{\ga}10^{23}{\rm W\,Hz}^{-1}$. \citet{yun} thus model the far-infrared LF as a sum of two Schechter functions, one for normal field galaxies, and a second to model the high-luminosity excess composed of a starburst/ULIRG population in which the SFRs (and hence FIR luminosities) are temporarily boosted by an order of magnitude. 

The apparent ability to fit the cluster far-IR LFs by a single Schechter function, could suggest that this second starburst component is missing from clusters. However, we note that the density of such luminous infrared galaxies in the field is ${\la}10^{-5}{\rm Mpc}^{-3}$, and given our survey volume of just ${\sim}2\,400\,{\rm Mpc}^{3}$ we would expect to find ${<<}1$ such galaxy in the SSC. 

\section{Summary}

We have presented new panoramic {\em Spitzer}/MIPS mid- and far-infrared (MIR/FIR) and {\em GALEX} ultraviolet imaging of the most massive and dynamically active system in the local Universe, the Shapley supercluster at $z{=}0.048$, covering the five clusters which make up the supercluster core. Using existing spectroscopic data from 814 confirmed supercluster members, we have produced the largest complete census of star-formation (both obscured and unobscured) in local cluster galaxies to date, extending down to ${\rm SFRs}{\sim}0.$02--0.$05\,{\rm M}_{\odot}{\rm yr}^{-1}$, providing a local benchmark for comparison to ongoing and future studies of cluster galaxies at higher redshifts with {\em Spitzer} and {\em Herschel}.
 From the {\em GALEX} data we produced near-ultraviolet (NUV) and far-ultraviolet (FUV) luminosity functions, which were found to have steeper faint-end slopes ($\alpha{=}{-}1.505$) than the local field population at the ${\sim}2\sigma$ level, due largely to the contribution of massive, quiescent galaxies at $M_{FUV}{\ga}{-}16$. Using the {\em Spitzer}/MIPS 24$\mu$m imaging, we determined the infrared galaxy LF of the Shapley supercluster core, finding it well described by a single Schechter function with $\log (L_{IR}^{*}/L_{\odot}){=}10.52_{-0.08}^{+0.06}$ and $\alpha(IR){=}{-}1.49{\pm}0.04$, fully consistent with those obtained for the Coma cluster and Abell 3266. We also presented the first 70$\mu$m galaxy luminosity function of a local cluster with {\em Spitzer}, finding it to be consistent with that obtained at 24$\mu$m. The shapes of the 24$\mu$m and 70$\mu$m galaxy luminosity functions were also found to be indistinguishable from those of the local field population. This apparent lack of environmental dependence for the shape of the FIR luminosity function suggests that the bulk of the star-forming galaxies that make up the observed cluster infrared LF have been recently accreted from the field and have yet to have their star formation activity significantly affected by the cluster environment. Combining the {\em GALEX} FUV and {\em Spitzer} 24$\mu$m photometry we estimate a global SFR of $327^{+102}_{-60}\,{\rm M}_{\odot}{\rm yr}^{-1}$ over the whole supercluster core, of which just ${\sim}2$0 per cent is visible directly in the ultraviolet continuum and ${\sim}8$0 per cent of the energy from star formation is reprocessed by dust and emitted in the infrared.

\section*{Acknowledgements}

This work was carried out in the framework of the collaboration of the FP7-PEOPLE-IRSES-2008 project ACCESS. CPH, RJS and GPS acknowledge financial support from STFC.  GPS acknowledges support from the Royal Society.  
This research made use of Tiny Tim/Spitzer, developed by John Krist for the Spitzer Science Center. The Center is managed by the California Institute of Technology under a contract with NASA. 
We thank the colleagues at the Spitzer Helpdesk for their kind cooperation.

\label{lastpage}
\end{document}